# Detecting hidden signs of diabetes in external eye photographs

**Authors**
Boris Babenko[1†], Akinori Mitani[1†], Ilana Traynis[2], Naho Kitade[1], Preeti Singh[1], April Maa[3,4], Jorge Cuadros[5], Greg S. Corrado[1], Lily Peng[1], Dale R. Webster[1], Avinash Varadarajan[1‡], Naama Hammel[1‡], Yun Liu[1‡]

**Affiliations**
[1]Google Health, Palo Alto, CA, USA
[2]Work done at Google Health via Advanced Clinical, Deerfield, IL, USA
[3]Department of Ophthalmology, Emory University School of Medicine, Atlanta, GA, USA
[4]Regional Telehealth Services, Technology-based Eye Care Services (TECS) division, Veterans Integrated Service Network (VISN) 7, Atlanta, GA, USA
[5]EyePACS Inc, Santa Cruz, CA, USA

[†]Equal contribution
[‡]Equal contribution

Correspondence: nhammel@google.com, liuyun@google.com

# Abstract

Diabetes-related retinal conditions can be detected by using a fundoscope or fundus camera to examine the posterior part of the eye. By contrast, examining or imaging the anterior part of the eye can reveal conditions affecting the front of the eye, such as changes to the eyelids, cornea, or crystalline lens. In this work, we studied whether external photographs of the front of the eye can reveal insights into both diabetic retinal diseases and blood glucose control. We developed a deep learning system (DLS) using external eye photographs of 145,832 patients with diabetes from 301 diabetic retinopathy (DR) screening sites in one US state, and evaluated the DLS on three validation sets containing images from 198 sites in 18 other US states. In validation set A (n=27,415 patients, all undilated), the DLS detected poor blood glucose control (HbA1c > 9%) with an area under receiver operating characteristic curve (AUC) of 70.2% (95%CI 69.4-70.9); moderate-or-worse DR with an AUC of 75.3% (95%CI 74.4-76.2); diabetic macular edema with an AUC of 78.0% (95%CI 76.4-79.6); and vision-threatening DR with an AUC of 79.4% (95%CI 78.1-80.8). For all 4 prediction tasks, the DLS's AUC was higher (p<0.001) than using available self-reported baseline characteristics (age, sex, race/ethnicity, years with diabetes). In terms of positive predictive value, the top 5% of patients based on DLS-predicted likelihood had a 67% chance of having HbA1c > 9%, and a 20% chance of having vision threatening diabetic retinopathy that needed ophthalmology followup (vs. 54% and 14%, respectively for baseline characteristics). Similarly, the odds ratio per standard deviation increase in the DLS prediction was 2.0 for HbA1c > 9% and 1.6 for vision threatening diabetic retinopathy after adjusting for baseline characteristics (p<0.001 for both). The results generalized to patients with dilated pupils in validation set B (n=5,058 patients) and to patients at a different screening service (validation set C, n=10,402 patients). Our results indicate that external eye photographs contain information useful for healthcare providers managing patients with diabetes, and may help prioritize patients for in-person screening. Further work is needed to validate these findings on different devices (e.g., computer web cameras and front-facing smartphone cameras) and patient populations (those without diabetes) to evaluate its utility for remote diagnosis and management.

# Introduction

Disease diagnosis and management often requires specialized equipment and trained medical professionals to interpret the findings. For example, diabetic retinopathy (DR) screening programs use either a fundoscope or a fundus camera to examine the posterior part of the eye (i.e., the retinal fundus, Figure 1A). Such examinations can reveal diabetes-related blood vessel compromise, such as microaneurysms. More recently, machine-learning based assessments of retinal fundus photographs has been shown to detect signs of cardiovascular risk,[1,2] anemia,[3] chronic kidney disease,[4] and other systemic parameters.[5] Despite the expanding diagnostic information that can be obtained from fundus photographs, the burden of costly specialized cameras, skilled imaging technicians, and oftentimes mydriatic eye drops to dilate the patient's' pupils, limits the use of these diagnostic techniques to eye clinics or primary care facilities with specialized equipment.

By contrast, the anterior of the eye can be readily imaged to produce external eye photographs (Figure 1A) without specialized equipment or mydriatic eye drops. Indeed, external eye photographs have an associated Current Procedural Terminology code (92285), and are sometimes used to document progression of anterior eye conditions such as eyelid abnormalities, corneal ulcers, and cataracts. Interestingly, the anterior part of the eye is also known to manifest signs of disease beyond that affecting the front of the eye. For example, hypertension can cause recurrent subconjunctival hemorrhages (from broken blood vessels in white part of the eye),[6] and diabetes can affect pupil size[7,8] and conjunctival vessel caliber[9] and tortuosity[10].

In this work, we hypothesized that deep learning can more precisely detect disease presence from external eye photographs, and potentially extract information that can help healthcare providers who manage diabetic patients. To test this hypothesis, we leveraged de-identified data from DR screening programs where important parameters of diabetes and diabetic eye care were available: HbA1c (as a marker of glucose control), and diabetic retinal diseases: DR, diabetic macular edema (DME), and vision-threatening DR (VTDR, a combination of DR and DME). Our results indicate that external eye photographs can indeed reveal signs of systemic and retinal disease, and merits further study.

# Results

We developed a deep learning system (DLS) using external eye images taken using fundus cameras (Figure 1A) from 145,832 diabetic patients screened for diabetic retinopathy (DR) at 301 sites in California, U.S.A. (Figure 1B). To evaluate the DLS, we used three validation datasets of patients with diabetes (Table 1). The first validation set ("A") included 27,415 patients who underwent DR screening at 186 sites across 18 U.S. states without pupillary dilation (i.e., without use of mydriatic eye drops); the second validation set ("B") included 5,058 patients at 59 sites across 14 U.S. states with pupillary dilation. The third validation set ("C") included 10,402 patients who underwent DR screening at the Atlanta Veterans Affairs (VA) in Decatur, GA, as part of the diabetic teleretinal screening program. The prespecified primary analyses included 4 predictions: poor blood glucose control (defined as HbA1c > 9% by the Healthcare Effectiveness Data and Information Set (HEDIS[11]), moderate-or-worse ("moderate+") DR, DME, and VTDR. Cataract detection was a fifth prediction task used as a positive control.

## Disease detection performance

We first evaluated the DLS for its ability to detect HbA1c > 9%. Across the three validation sets, the DLS achieved an area under the receiver operating characteristic curve (AUC) of 70.2%, 73.4%, and 69.8% respectively (Figure 2A). For comparison, logistic regression models using baseline characteristics (age, sex, race/ethnicity, years with diabetes on validation sets A and B; age and sex on validation set C) achieved AUCs of 64.8%, 66.2%, and 65.3% respectively (p<0.001 for comparing the DLS with baseline characteristics alone in all three validation sets). The AUC trends for predicting other thresholds (HbA1c > 7% and > 8%) were similar and are presented in Supplementary Table 1. The positive predictive values (PPVs) for HbA1c > 9% exceeded 45% for the 5% of patients with the highest-risk predicted by the DLS, and the negative predictive values (NPVs) exceeded 65% for the lowest-risk 5% (Figure 1C, Supplementary Figures 1A and 2A).

Next, we evaluated the DLS for detecting diabetic retinal diseases: moderate+ DR, DME, and VTDR (Figures 2B-D). The AUCs for the three conditions were 75.3%, 78.0%, and 79.4% in validation set A. On dilated pupils, the AUCs were consistently 5-10% higher at 84.2%, 85.2%, and 87.1% in validation set B, and more similar at 76.7%, 75.8%, and 79.0% in validation set C (which was from a different patient population). Using only baseline characteristics resulted in AUCs that were 4-10% lower in validation sets A and B, and 10-20% lower in validation set C (p<0.001 for comparing the DLS with baseline characteristics alone for all comparisons). The trends for other DR thresholds ("mild+" and "severe+") were similar and are presented in Supplementary Table 1. The PPV and NPV for the various diabetic retinal diseases similarly indicated the ability to identify patients at either very high or very low likelihood of these conditions (Figure 1C and Supplementary Figures 1B-D and 2B-D).

## Adjusted analysis

To verify that the DLS predictions are not solely mediated by baseline characteristics as confounders, we examined the odds ratios of the DLS predictions adjusted for baseline characteristics (Supplementary Table 2). For all prediction tasks in all three validation datasets, DLS predictions (standardized to zero mean and unit variance for interpretability of coefficients) had statistically significant adjusted odds ratios (p<0.001). The adjusted odds ratios ranged from 1.4 to 2.0 among tasks for validation set A, from 2.0 to 3.1 for validation set B, and from 1.6 to 2.0 for validation set C.

## Explainability analysis

To better understand how the DLS was able to detect these diseases without looking inside the eye at the retina, we next conducted several explainability experiments. Predicting cataract presence acted as a positive control because it manifests as media opacity ("cloudiness" of the lens) and thus is generally visible from external eye images. The AUC of the DLS for predicting cataract was substantially higher than that for predicting high HbA1c or diabetic retinal diseases, at 86.7% and 93.4% for validation sets A and B (labels were not available for set C), similarly higher than using baseline characteristics (Figure 2E and Supplementary Table 1).

We first conducted ablation analysis, where portions of each image were removed, and another DLS trained on such images was evaluated (Methods).[3] The central region of the image was more important than the outer rim for all predictions, with the biggest delta being our positive control, cataract (Figure 3A-B).

Next, we conducted saliency analysis using several gradient-based methods (Methods). The HbA1c prediction was the most striking, with almost all of the attention on the nasal and temporal conjunctiva/sclera (i.e., where conjunctival vessels tend to be prominent, see Figure 4A-B). By contrast, all three diabetic retinal disease predictions (moderate+ DR, DME, VTDR) focused on both the pupil and the nasal and temporal iris and conjunctiva (Figure 4C-E). The cataract predictions focused on the pupil, in the middle of the image (Figure 4F). A quantitative summary of these trends is presented in Figure 4G-H and the results of additional saliency Methods and for other prediction tasks are presented in Supplementary Figure 3.

Because diabetes is known to affect pupil sizes,[7,12] and multiple attribution methods highlighted the pupil/iris region (Figure 5 and Supplementary Figure 3), we reasoned that the DLS could have learnt to perform these predictions via measuring pupil size. To test this hypothesis, we first trained a pupil/iris detection model using a subset of the development set (Methods), and used this model to estimate the pupil size in the validation sets (Supplementary Figure 4). To correct for differences in distance from the camera, we normalized the pupil size by representing the pupil radius as a fraction of the iris radius. The iris size is equivalent to the corneal diameter since the visible iris extends to the edge of the corneoscleral junction. Corneal diameter (or iris size here) is known to be relatively constant in the adult population.[13,14]

We first examined whether the baseline model performance changes if we include pupil size among the baseline characteristics (Supplementary Table 3). There was generally little to no improvement in AUC of the baseline characteristics model (approximately 1% with overlapping confidence intervals). The only exceptions were in validation set C, with 3-4% AUC improvements for moderate+ DR and VTDR. The improvement of the DLS over these augmented baseline characteristics models remained statistically significant in all cases ($p<0.001$). Similarly, in adjusted analysis including estimated pupil size, the external eye predictions retained statistically significant odds ratios (Supplementary Table 4).

## Sensitivity and subgroup analysis

Because these external eye images were taken with fundus cameras, we next sought to examine if lower-quality images would suffice. To simulate low-quality images, the input images were downsampled to lower resolutions for both training and evaluation (Methods). The DLS performance decreased slightly as the input size decreased to 75 pixels, and then substantially as the input size decreased below 35 pixels (Figure 3C-D).

Finally, we conducted extensive subgroup analysis across multiple variables: demographic information (age, sex, race/ethnicity), presence of cataract, HbA1c levels and pupil size (Supplementary Tables 5-9). Briefly, the DLS (and the baseline variables) remained predictive in all subgroups. The DLS consistently outperformed the baseline characteristics model, albeit with the difference not reaching statistical significance in some subgroups. The DLS's AUC dropped by more than 5% in some subgroups: Black for moderate+ DR; Black, >10 years with diabetes, and cataract for VTDR; and age > 50 years and small pupil for cataract. In all of these cases, the AUC of the baseline characteristics model decreased as well.

## Discussion

Our results show that external images of the eye enable detection of diabetes-related conditions (severity and type of diabetic retinal disease) that normally require specialized retinal photographs, as well as poor blood sugar control. We accomplished this through the development of a DLS, which generalized to diverse patient populations, different imaging protocols, and devices from independent clinics in multiple U.S. states. Our results were significantly better at these predictions than using demographic information and medical history (such as years with diabetes) alone, and remained significant after adjusting for multiple baseline characteristics and within numerous subgroups.

Our technique has implications for the large and rapidly growing population of patients with diabetes because it does not, in principle, require specialized equipment. Specifically, detection of diabetes-related retinal disease requires fundoscopy or the use of a fundus camera to examine the back of the eye through the pupil. This limits disease screening and detection exams to either eye clinics or store-and-forward teleretinal screening sites where fundus cameras are present -- both of which require in-person visits. Indeed, poor access and cost contributes, in part, to poor DR screening rates.[15] Similarly, a HbA1c measurement requires an in-person visit for an venous blood draw, which can be unpleasant for patients and have multiple potential side effects including bleeding, bruising, and nerve damage.[16] By contrast, our technique requires only a photograph of the front of the eye without pupil dilation via eye drops. Our experiments further show that even low-resolution images of 75 × 75 pixels (which is 1% of the resolution of a basic "720p" laptop webcam and 0.1% of the resolution of a standard 8-megapixel smartphone camera) results in adequate performance, suggesting that the resolution requirements for this technique can be easily met. While further work is needed to determine if there are additional requirements for lighting, image stabilization, lens quality, or sensor fidelity, we hope that disease detection techniques via external eye images can eventually be widely accessible to patients, whether in clinics, pharmacies, or even at home. Because access to and utilization of healthcare is associated with poor diabetic parameters[17], we hope that the availability of easily-accessible techniques such as ours can improve adherence to diabetic interventions which we know to be cost saving or cost effective.[18]

The specific use cases for easy identification and monitoring of high-risk diabetic patients are manifold. First, detecting diabetic patients who have difficulty controlling their blood glucose (HbA1c > 9%) may help to reveal which patients are in need of further counseling, additional diabetic resources and medication changes. In our analysis, when the top 5% of patients with the highest predicted likelihood were examined, 50-75% may truly have HbA1c > 9%. Similarly, if the findings (e.g., HbA1c > 7%) can be shown in future studies to generalize to patients without diagnosed diabetes, identification of these potentially asymptomatic patients at risk for early or mild diabetes can help determine which patients may benefit from a confirmatory blood test and early interventions such as lifestyle counseling or medications. Second, identification of patients at risk for diabetic retinal disease can determine which patients may benefit from ophthalmology followup and targeted treatment to avoid diabetes-associated vision loss. In our analysis, if the top 5% of patients with the highest predicted likelihood of various diabetic retinal diseases were examined via fundus photographs, 20-65% could have vision-threatening diabetic retinal disease and 20-90% could have moderate-or-worse diabetic retinal disease that warrant ophthalmology followup. Importantly, since external eye photos do not require specialized equipment, DR screening using this method may help to close the access gap that has contributed to the lack of appropriate diabetic eye disease screening in approximately half of patients with diabetes. Conversely, patients at very low risk of developing diabetic retinal disease could potentially be screened less frequently. Remote identification of patients who would benefit from in-person specialized eye care and treatment, could potentially allow for earlier diagnosis, treatment, and better outcomes in these high-risk individuals. In addition, patients who are found to be at low risk of diabetic retinal disease could avoid the time and cost of missing work and traveling to a specialized eye clinic for an in-person examination. Lastly, minimizing in-person visits to health clinics has become exceedingly important in preventing disease spread. During the current COVID-19 pandemic, many countries are actively encouraging "social" (physical) distancing and deferment of non-urgent medical visits, including decreased visits for diabetes.[19] Similar to how phone cameras[20,21] can help identify digital biomarkers of diabetes[22], external eye photo-based screening could allow for remote evaluation, potentially from home, and with common equipment, greatly minimizing risk to the individual and others.

Scientifically, our discovery that predictions about diabetic disease states could be derived from external eye photography is surprising since such images are primarily used to identify and monitor anterior eye conditions, such as eyelid and conjunctival malignancies, corneal infections, and cataracts. Although there are numerous studies noting conjunctival vessel changes (fewer, wider, and less tortuous conjunctival vessels)[9,10,23–25] associated with duration of diabetes[26,27] and severity of diabetic retinopathy[28,29], to our knowledge there have been no large studies linking HbA1c or diabetic macular edema to conjunctival vessel changes in diabetes. Furthermore, conjunctival vessel assessment for signs of diabetes is not a common clinical practice due to the subjectivity and time consuming nature of such an evaluation and the option of a more accurate and easier test for the clinician (i.e. HbA1c). We verified that these surprising results were reproducible and not an artifact of a single dataset or site, via broad geographical validation across 18 U.S. states.

Our evidence provides several hints as to how these predictions are possible. First, the ablation analysis indicates the center of the image (pupil/lens, iris/cornea, conjunctiva/sclera) are substantially more important than the image periphery (e.g., eyelids) for all predictions (Figure 3). Second, the saliency analysis similarly indicates that the DLS is most influenced by areas near the center of the image. These included the pupil and the corneoscleral junction for diabetic retinal disease, and at both the nasal and temporal conjunctiva (where conjunctival blood vessels are often most prominent), for blood sugar control (i.e., HbA1c) (Figure 4). In both analyses, the positive control (cataract, which is visible at the pupil/lens) provides a useful baseline attribution map for a prediction that is expected to focus exclusively on the pupil/lens. Together, these data suggest that the DLS is leveraging both information in the pupil region, such as from the light reflecting from the retina ("red reflex"), and information outside of the pupil/iris region, such as in the conjunctival vessels. Finally, the DLS's performance for HbA1c trended upwards for higher HbA1c cutoffs (i.e., AUC: 67% for >7% vs 69% for >8% vs 70% for >9%), compared to a "flat" trend of 65% for baseline characteristics. This "dose-dependent" trend cannot be explained by increased training data (because the number of examples of >9% is strictly less than >8%), and suggests that the features used by the DLS were related to the extent of elevated glucose levels. Better scientific understanding of these predictions via systematic falsifying of hypotheses will need to be addressed in future work.

This work establishes that signs of systemic disease, specifically diabetes, can be identified from external eye images. Looking beyond diabetes, these findings raise the tantalizing prospect that such external eye photographs may contain additional useful signals, both familiar and novel, related to other systemic conditions. Clinically, numerous external or anterior segment findings are correlated with or the result of underlying systemic disease. Obstructive sleep apnea is associated with floppy eyelid syndrome (easily everted eyelids) and resultant papillary conjunctivitis[30,31]. Thyroid disease can manifest specific ocular signs such as upper eyelid retraction, conjunctival injection (redness) and chemosis (swelling), and caruncular edema (swelling of small, pink nodule at the inner corner of the eye)[32,33]. Elevated cholesterol levels and atherosclerosis have been linked with xanthelasmas (yellowish deposit under the eyelid skin)[34], which are predictive of adverse cardiac outcomes[35]. Systemic hypertension, another high-morbidity disease, has been associated with conjunctival microangiopathy and correlates with time since hypertension diagnosis.[36] Beyond eyelid and conjunctival changes, deposition of calcium or uric acid in the cornea may signify derangements related to hyperparathyroidism, chronic renal failure, and gout[37–39]. These manifestations could be readily captured with external eye photography, which document the eyelids, conjunctiva, and cornea, and further work examining these disease populations are needed to assess additional systemic disease prediction models. Similar to how investigations into manifestations of systemic disease in the retina has been dubbed "oculomics"[40], such analyses on external ocular images could be termed "exoculomics."

Our study contains limitations. First, we have limited data with which to understand if smartphone or webcam images are sufficient because all images were taken by healthcare professionals using a table-top fundus camera. Though the resolution and actual image sensor (for example Canon EOS series DSLRs) were comparable to consumer-grade cameras, these clinical-grade cameras provided chin stabilization that may have reduced motion blur, good lighting to reduce occlusion of features from shadows, and greater magnification[41]. Similarly, despite only semi-standardized protocols, clinical professionals may have provided instructions to standardize gaze and improve image quality, and the clinical environment may have been darkened to enlarge pupil size. Additional work is needed to evaluate if the DLS generalizes to other consumer-grade cameras and settings without modification, or if additional data collection is needed. Second, several potentially relevant baseline characteristics such as body-mass index[42] were not available and hence could not be included in the adjusted analyses or subgroup analyses. Lastly, our study included only patients with diabetes in DR screening programs, and generalization to non-diabetic patients will need to be further evaluated.

In conclusion, we have demonstrated that external eye images can be used to detect the presence of several diabetes-related conditions such as poor blood sugar control and various diabetic retinal diseases. Further study is warranted to evaluate if such a tool can be used in a home, pharmacy, or primary care setting to improve disease screening and help with management of diabetes.

## Methods

### Datasets

This work utilized three teleretinal diabetic screening datasets in the U.S. (Table 1). The first two datasets were the California and non-California cohorts from EyePACS, a teleretinal screening service in the U.S.[43]. In the EyePACS datasets, patients presented to sites such as primary care clinics for diabetic retinal disease screening. Visits with unknown sites were excluded; and in addition, patients associated only with sites with unknown states were excluded. We leveraged data from California (the state with the most visits in EyePACS) for development, and the remaining U.S. states for validation purposes. Within the development dataset, data from Los Angeles county were further excluded as held-out data for another project. Validation set A consisted of all non-California cases without pupil dilation, and validation set B consisted of all non-California cases with pupil dilation. The third validation dataset ("C") was from the Atlanta Veterans Affairs (VA) Eye Clinic diabetic teleretinal screening program, which served multiple community-based outpatient clinics (CBOCs) in the greater Atlanta area. Most visits in validation set C involved pupil dilation. Ethics review and Institutional Review Board exemption for this retrospective study on de-identified data were obtained via Advarra Review Institutional Review Board.

## Imaging protocol

As part of the standard imaging protocol for diabetic retinopathy screening at these sites, patients had photographs taken of the external eye for evaluation of the anterior segment (along with the retinal fundus photograph). At sites served by EyePACS, cameras included Canon (CR1 and CR2), Topcon (NW200 and NW400), Zeiss Visucam, Optovue iCam, and Centervue DRS. Based on available image Exif metadata, images were digitized using Canon Electro-Optical System (EOS) single-lens reflex cameras (SLR). The image protocol involved first positioning patients comfortably in front of the camera with their chin resting on the chin rest and their forehead an inch away from the forehead brace. The operator then took external images of the right and then the left eye, ensuring image clarity, focus, and visibility of the entire eye, including the iris, pupil, and sclera.[44] At sites served by the VA diabetic teleretinal screening program, the cameras included Topcon NW8 and Topcon NW400. The acquisition protocol involved slightly distancing the patient from the fundus camera to capture a clear view of the external eye; further alignment was entrusted to the operator. The external eye photographs were intended to document findings such as cataracts and eyelid lesions.

## Labels

Diabetic retinal disease in EyePACS (development set and validation sets A and B) was graded by EyePACS-certified graders using a modified Early Treatment Diabetic Retinopathy Study (ETDRS) grading protocol.[45,46] Graders evaluated three-field retinal fundus photographs (nasal, primary, and temporal), and the absence or presence of lesions were mapped[47] to 5 International Clinical Disease Severity Scale (ICDR) DR levels: no DR, mild, moderate, severe, and proliferative DR and DME. VTDR was defined as severe-or-worse DR, or DME. Diabetic retinal disease in the VA dataset (validation set C) was graded by ophthalmologists at the VA. Baseline characteristics including demographic information were patient-reported and provided by each site. HbA1c was extracted by each site from their medical record and could be from a different visit. Cataract presence in the EyePACS cases was indicated by the same graders.

To better understand if pupil size was associated with our results, we obtained segmentations of both pupil and iris size. To do so, 12 ophthalmologist graders evaluated each external eye image. If the pupil and iris were distinct enough to delineate the borders accurately, the graders drew ellipses around both the iris and pupil (Supplementary Figure 5). This was performed for a subset of 4000 randomly selected images in the development set, another 500 images in validation sets A and B combined, and 500 images in validation set C. We then trained a model to segment the iris and pupil using the labeled development set images, and evaluated the accuracy in the validation sets (see "Pupil/iris Segmentation Model" and Supplementary Figure 4 in Supplementary Information). Our pupil size and image quality subgroup analyses were based on running this model across all images.

## Deep learning system development

To develop the external eye DLS, we split the development dataset into a training and tuning split in a 7:1 ratio, while ensuring each site was only in one split. All visits for all patients were used for training or tuning. A multi-task learning approach was used to train a single network to predict all tasks (i.e., one "head" per task). Specifically, all heads adopted a classification setup (with cross entropy loss) to ensure all losses were in comparable units. HbA1c was trained as 3 binary classification heads with thresholds at 7%, 8%, and 9%; DR was discretized into the 5 ICDR categories; DME was binary (present/absent); VTDR was binary (present/absent); and cataract was binary (present/absent). Some additional "auxiliary" heads were included: age discretized into 7 categories: (0, 30], (30, 40], (40, 50], (50, 60], (70, 80] and (80, 90] years; race/ethnicity categorized into Hispanic, White, Black, Asian and Pacific Islander, Native American and others; sex represented as male and female; and years with diabetes discretized into 5 categories: (0, 1.5], (1.5, 5.0], (5.0, 10.5], (10.5, 15.5], (15.5, inf] years. Because not every label was available for all of these heads, the losses were propagated only for the relevant heads for each example.

We trained 5 models based on the Inception-v3[48] architecture using hyperparameters summarized in Supplementary Table 10, and the predicted likelihood was averaged unless otherwise noted. Within the hyperparameter search space detailed in Supplementary Table 10, 100 random trials were conducted and the top 5 models were selected. During each training run, the tune-set AUC for HbA1c > 7% was the first to decline after reaching the peak, so early stopping was based on the tune-set AUC of this task.

## Deep learning system evaluation

For evaluation, we selected a single visit per patient at random. This resulted in 27,415 visits for validation set A, 5,058 for set B, and 10,402 for set C. We preprocessed the external eye photographs in the same manner as in previous work for color fundus photographs[1,49]. Some of the prediction targets (described below) are on a per-visit basis (e.g., HbA1c has a single value per visit per patient), and so the predictions of both eyes were averaged for evaluation purposes. For prediction targets that are specifically for each eye (e.g., eye diseases that affect each eye individually), we randomly selected one eye per visit during evaluation. The exact numbers of data points used for each prediction task are presented in Supplementary Table 1.

For validation set A and B, visits without dilation status information were excluded (9% of the total visits). A small number of patients had both visits with dilation and visits without dilation, and were included in both datasets (n=497).

## Baseline models for comparisons

Models leveraging baseline characteristics were used as a baseline, and were trained using logistic regression with class-balanced weighting and the default L2 regularization (C=1.0) in the scikit-learn Python library. For validation sets A and B, the input baseline characteristics were self-reported variables: age, sex, race/ethnicity, and years with diabetes. This baseline model was trained using the entire training dataset because using only undilated or dilated cases did not improve tune set performance. This also kept the development process of the baseline model consistent with that of the DLS. For validation set C, because race/ethnicity and years with diabetes were not available, only age and sex were used as input variables. This baseline model was trained using validation set C itself due to the large differences in patient population (mostly male and older, and enriched for black and white patients), and differences in available baseline characteristics. This overestimates the baseline model's performance and underestimates the improvement (delta) of our DLS.

## Statistical analysis

We evaluated all results using the AUC, expressed as percentage (i.e., 0-100%, where 50% is random performance). To evaluate the superiority of DLS predictions compared to the baseline model predictions in each dataset, we used the DeLong method[50]. The superiority analysis on 4 tasks (HbA1c > 9%, moderate+ DR, DME, VTDR) were prespecified and documented as primary analyses before applying the DLS to the validation sets. Alpha was adjusted using the Bonferroni method for multiple comparison correction (α=0.025 for one-sided superiority test, divided by 4 tasks=0.00625).

Prespecified secondary analyses include additional evaluation metrics (sensitivity, specificity, PPV, and NPV), additional related prediction tasks (cataract, HbA1c with different thresholds: mild+ and severe+ DR), subgroup analysis (presence of cataract, pupil size, HbA1c stratification, demographic groups based on age, sex, race/ethnicity), adjusted analysis, explainability analysis, and sensitivity analysis for image resolution, dilation status, and temporal difference of HbA1c lab and image. Adjusted analyses leveraged logistic regression models fit using the statsmodels library v0.12.1. For subgroup analysis, when a patient had multiple visits that satisfied a filtering criteria, one visit was chosen randomly.

## Ablation analysis

We evaluated the dependence of DLS performance on the visibility of different regions of the image. Because both the images and ocular anatomy are "circular" (i.e., roughly radially symmetric), and most pupils are centered, we conducted this "image visibility" analysis based on concentric circles. Two types of masking were examined: keeping only the center visible and keeping the peripheral rim visible. At various degrees of masking, the two types of masking were compared while controlling for the amount of visible pixels. For each condition, the same mask was applied during both training and evaluation.

## Saliency analysis

From validation set A, 100 positive images with the highest predicted likelihoods were selected for saliency analysis for each task. Images were manually inspected to ensure correct vertical orientation and to orient the nasal aspect on the right. Three saliency methods were applied: Grad-CAM[51], guided backprop[52], and integrated gradient[53], as implemented in the People+AI Research saliency library[54].

## Image resolution analysis

When simulating the low-resolution images, images were first downsampled to the specified resolution using the area-based method (tf.image.resize with method=AREA). Next, to ensure the DLS's input size (and thus the network's number of parameters and capacity) remained the same for a fair comparison, we upsampled each image back to the original resolution (by bilinear interpolation with antialias using tf.image.resize with antialias=True). Downsampling and upsampling methods were chosen to produce most visually appropriate images, blinded to the actual DLS results.

## Data availability

This study utilized de-identified data from EyePACS Inc. and the teleretinal diabetes screening program at the Atlanta Veterans Affairs. Interested researchers should contact J.C. (jcuadros@eyepacs.com) to inquire about access to EyePACS data and approach the Office of Research and Development at https://www.research.va.gov/resources/ORD_Admin/ord_contacts.cfm to inquire about access to VA data.

## Acknowledgments

This work was funded by Google LLC. The authors would like to acknowledge Huy Doan, Quang Duong, Roy Lee, and the Google Health team for software infrastructure support and data collection. We also thank Tiffany Guo, Mike McConnell, Michael Howell, and Sam Kavusi for their feedback on the manuscript. Last but not least, gratitude goes to the graders who labeled data for the pupil segmentation model.

## Competing interests

B.B., Akinori Mitani, N.H., G.S.C., L.H.P., D.R.W., A.V., N.H., and Y.L. are employees of Google LLC, own Alphabet stock, and are co-inventors on patents (in various stages) for machine learning using medical images. I.T. is a consultant of Google LLC. J.C. is the CEO of EyePACS Inc.

## Author contributions

B.B., Akinori Mitani, and N.K. conducted the machine learning model development, experiments, and statistical analysis with input and guidance from A.V., N.H., and Y.L.. B.B., Akinori Mitani, I.T., N.H., and Y.L.. designed the study and pre-specified the statistical analysis. I.T., N.K. and P.S. developed guidelines and managed data collection for the pupil / iris segmentation model. J.C. and April Maa managed data collection and associated approvals. G.S.C., L.H.P., and D.R.W. obtained funding for data collection and analysis, supervised the study, and provided strategic guidance. B.B., Akinori Mitani., I.T., N.H., and Y.L. prepared the manuscript with input from all authors. B.B. and Akinori Mitani contributed equally as co-first authors; A.V., N.H., and Y.L. contributed equally as co-last authors.

## Code availability

The deep learning framework (TensorFlow) used in this study is available at https://www.tensorflow.org; the neural network architecture is available from https://github.com/tensorflow/models/blob/master/research/slim/nets/inception_v3.py; and an ImageNet pretrained checkpoint is available from https://github.com/tensorflow/models/tree/master/research/slim.

# Figures

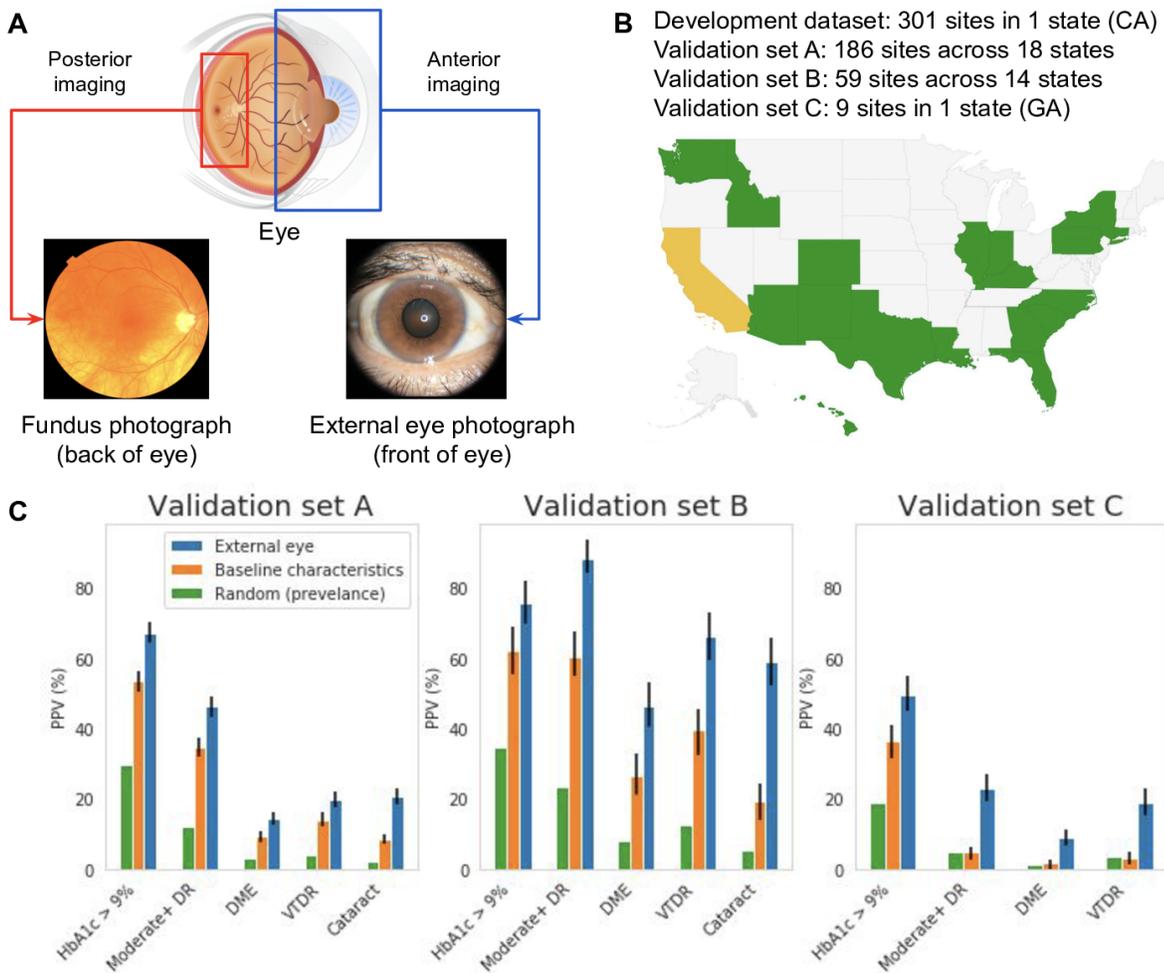

**Figure 1. Our study focuses on extracting insights from external photographs of the front of the eye.** (**A**) Diabetes-related complications can be diagnosed by using specialized cameras to take fundus photographs, which visualize the posterior segment of the eye. By contrast, anterior imaging using a standard consumer-grade camera can reveal conditions affecting the eyelids, conjunctiva, cornea, and lens. In this work, we show that external photographs of the eye can offer insights into both diabetic retinal disease and detect poor blood sugar control. These images are shown for illustrative purposes only and do not necessarily belong to the same subject. (**B**) Our external eye deep learning system (DLS) was developed on data from California (CA, in yellow) and evaluated on data from 18 other U.S. states (in green). (**C**) The positive predictive value (PPV) is shown for identifying patients at high risk of poor sugar control (HbA1c > 9%), moderate-or-worse diabetic retinopathy (moderate+ DR), diabetic macular edema (DME), vision-threatening DR (VTDR), and a positive control: cataract. The thresholds used here are based on the 5% of patients with the highest predicted likelihood. Baseline characteristics models for validation sets A and B include self-reported age, sex, race/ethnicity and years with diabetes and were trained on the training dataset. The baseline characteristics models for validation set C use self-reported age and sex and were trained directly on validation set C due to large differences in patient population compared to the development set. Error bars indicate 95% bootstrap confidence intervals. A graphic visualization of the PPV over multiple thresholds is in Supplementary Figure 1.

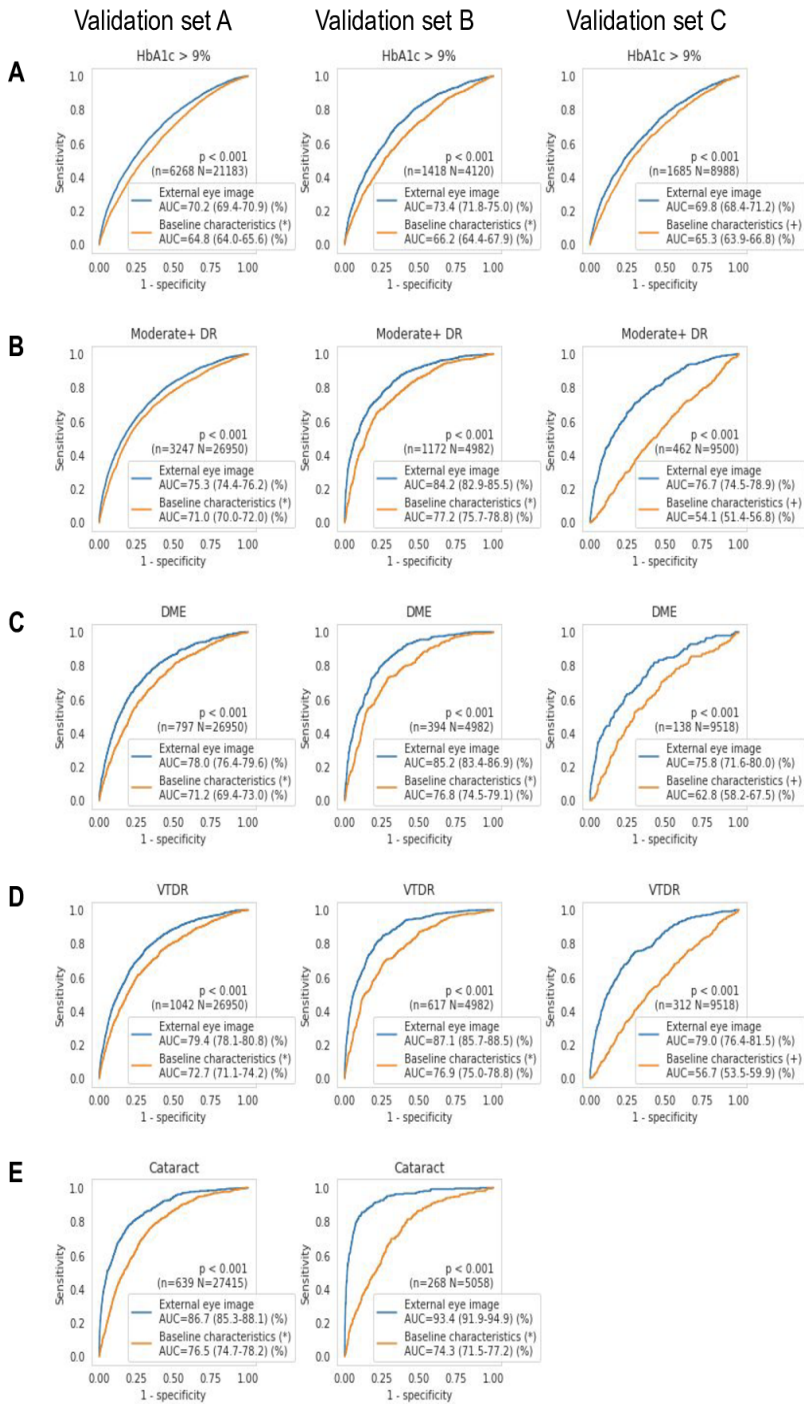

**Figure 2. Receiver operating characteristic curves (ROCs) for various predictions using external eye images: (A) poor sugar control (HbA1c > 9%), (B) moderate-or-worse diabetic retinopathy (moderate+ DR), (C) diabetic macular edema (DME), (D) vision-threatening DR (VTDR), and (E) a positive control: cataract.** Sample sizes ("N" for the number of cases and "n" for the number of positive cases), area under ROC (AUCs), and the p-value for the difference are indicated in the legend of each panel. The last panel is empty because "cataract" labels were not available for validation set C. (*) Baseline characteristics models for validation sets A and B include self-reported age, sex, race/ethnicity and years with diabetes and were trained on the training dataset. (+) The baseline characteristics models for validation set C use self-reported age and sex and were trained directly on validation set C due to large differences in patient population compared to the development set.

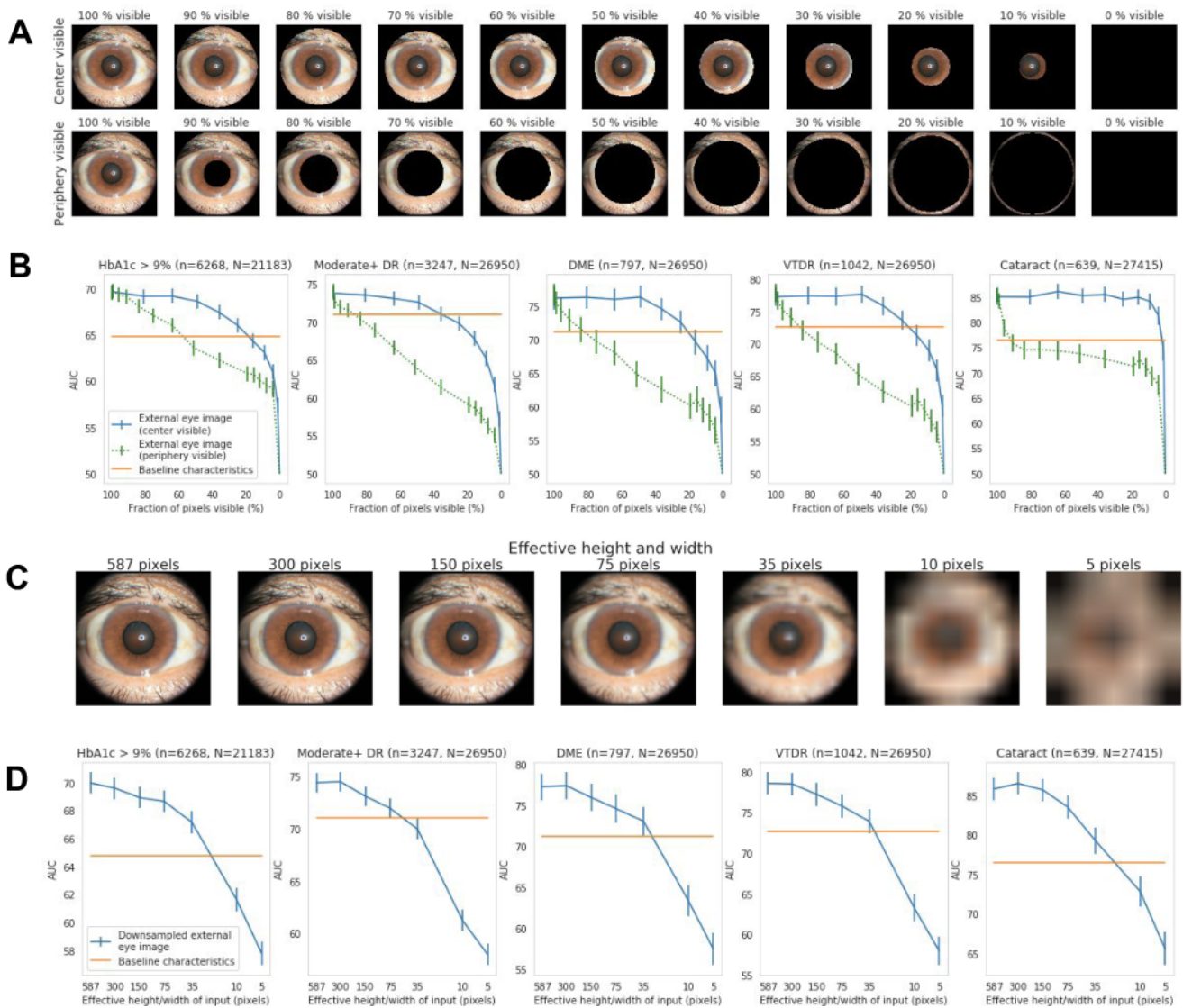

**Figure 3. Importance of different regions of the image and impact of image resolution. (A)** Sample images with different parts of the image visible. Top: only a circle centered around the center is visible; bottom: only the peripheral rim is visible. **(B)** Area under receiver operating characteristic curves (AUCs) as a function of the fraction of input pixels visible when the DLS sees only the center (blue solid curve) or peripheral regions (green dotted curve). A large delta between the blue and orange curves (e.g., for cataract) indicates most of the information is in the region corresponding to the blue line. **(C)** Sample images of different resolutions. Images were downsampled to a certain size (e.g., "5" indicates 5×5 pixels across), and then upsampled to the same original input size of 587×587 pixels to ensure the network has the exact same capacity (Methods). **(D)** AUCs corresponding to different input image resolutions.

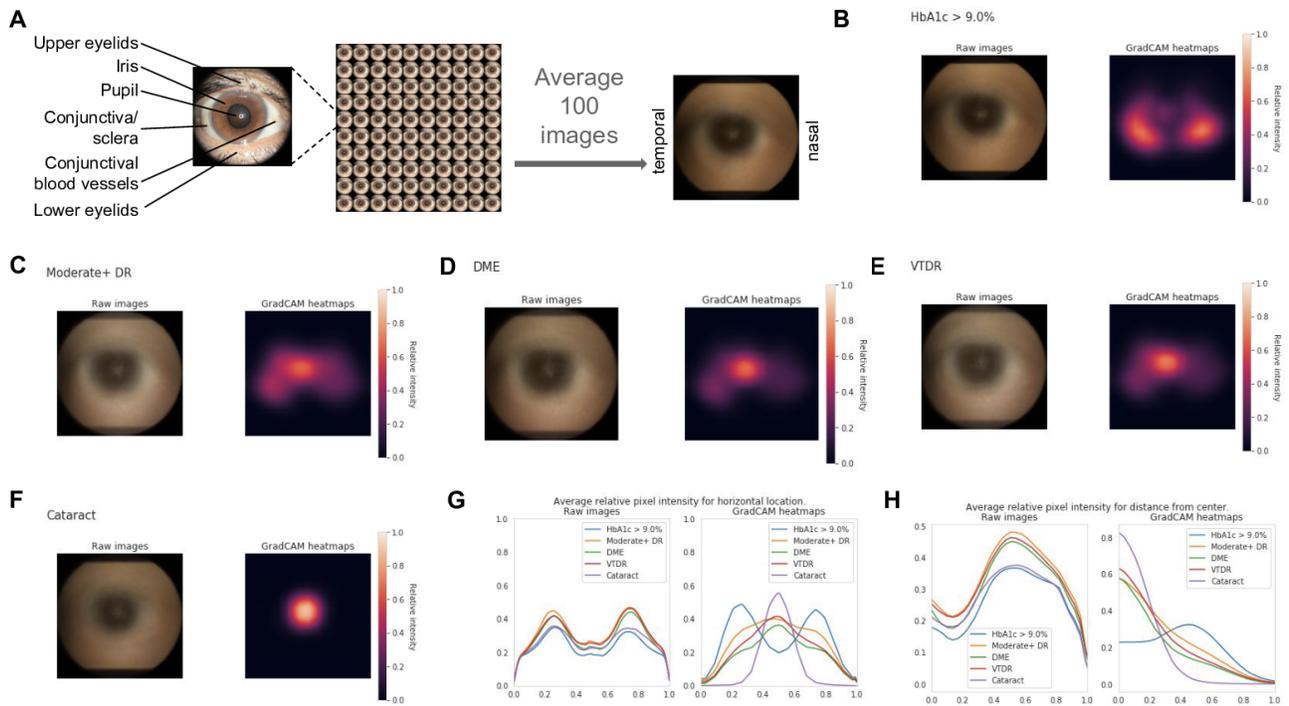

**Figure 4. Qualitative and quantitative saliency analysis illustrating the influence of various regions of the image towards the prediction. (A)** Overview of anterior eye anatomy; the illustrations in panels B-F are the result of averaging images or saliency maps from 100 eyes, with laterality manually adjusted via horizontal flips to have the nasal side on the right (Methods). **(B-F)** Saliency analysis with the original averaged image on the left; saliency from GradCAM on the right. **(G-H)** Quantitative comparison of the intensity of the pixels along the x-axis and in terms of distance from the center, respectively.

## Tables

**Table 1. Dataset characteristics**.

| Datasets | | Development set | | Validation sets | | |
|---|---|---|---|---|---|---|
| | | Training set | Tuning set | Validation set A | Validation set B | Validation set C |
| Source | | EyePACS (CA) | | EyePACS (non-CA) | | VA |
| No. of US states | | 1 (CA) | 1 (CA) | 18 (non-CA) | 14 (non-CA) | 1 (GA) |
| No. of sites | | 277* | 54* | 186* | 59* | 9 |
| No. patients | | 126,066 | 19,766 | 27,415 | 5,058 | 10,402 |
| No. visits** | | 159,269 | 23,095 | 27,415 | 5,058 | 10,402 |
| No. of images for diabetic retinal disease prediction** | | 290,642 | 41,928 | 27,415 | 5,058 | 10,402 |
| No. of images for HbA1c prediction** | | 290,642 | 41,928 | 53,861 | 9,853 | 19,763 |
| Dilation status (%) | No | 137,710 (47%) | 21,063 (50%) | 27,415 (100%) | 0 (0%) | 150 (1%) |
| | Yes | 100,929 (35%) | 9,678 (23%) | 0 (0%) | 5,058 (100%) | 9,788 (94%) |
| | Unknown | 52,003 (18%) | 11,187 (27%) | 0 (0%) | 0 (0%) | 464 (4%) |
| Age (years, mean ± std. dev.) | | 54 ± 11 | 54 ± 11 | 54 ± 12 | 53 ± 11 | 63 ± 10 |
| Self-reported sex (%) | Female | 72,880 (58%) | 11,655 (59%) | 14,525 (53%) | 3,004 (59%) | 478 (5%) |
| | Male | 53,186 (42%) | 8,111 (41%) | 12,890 (47%) | 2,054 (41%) | 9,924 (95%) |
| Race / ethnicity (%) | Hispanic | 97,300 (77%) | 15,577 (79%) | 11,393 (42%) | 4,152 (82%) | *** |
| | White | 11,459 (9%) | 1,609 (8%) | 7,621 (28%) | 374 (7%) | 45%*** |
| | Black | 4,991 (4%) | 1,074 (5%) | 4,140 (15%) | 398 (8%) | 49%*** |
| | Asian / Pacific islander | 7,954 (6%) | 1,091 (6%) | 2,949 (11%) | 62 (1%) | <1% |
| | Native American | 1,669 (1%) | 83 (0%) | 411 (1%) | 37 (1%) | <1% |
| | Other | 2,693 (2%) | 332 (2%) | 901 (3%) | 35 (1%) | *** |
| Years with diabetes (years, median (inter quartile range)) | | 5 (2 - 13) | 8 (2 - 13) | 5 (2 - 8) | 8 (2 - 13) | N/A |

* 30 sites overlapped between the training set and tuning set because patients who visited both training set sites and tuning set sites were randomly assigned to either set. 56 sites overlapped between validation sets A and B because they included both dilated and non-dilated patients. There was no patient-level overlap between the development set and the validation sets.

** In the development set, all visits, eyes, and images were used. In each validation set, 1 random visit was selected per patient. Though all evaluation was performed on the patient level, for eye disease evaluation 1 random eye per patient was selected, and for HbA1c evaluation images of both eyes were used (and the DLS predictions averaged).

***Case-level information not available; these reflect a cohort-level estimate, and is consistent with the predominantly black / white population in prior studies[55]
Abbreviations: CA = California, VA = Veterans Affairs, HbA1c = glycated hemoglobin.

# References

1.  Poplin, R. *et al.* Prediction of cardiovascular risk factors from retinal fundus photographs via deep learning. *Nature Biomedical Engineering* **2**, 158–164 (2018).

2.  Cheung, C. Y. *et al.* A deep-learning system for the assessment of cardiovascular disease risk via the measurement of retinal-vessel calibre. *Nat Biomed Eng* (2020) doi:10.1038/s41551-020-00626-4.

3.  Mitani, A. *et al.* Detection of anaemia from retinal fundus images via deep learning. *Nat Biomed Eng* **4**, 18–27 (2020).

4.  Sabanayagam, C. *et al.* A deep learning algorithm to detect chronic kidney disease from retinal photographs in community-based populations. *The Lancet Digital Health* **2**, e295–e302 (2020).

5.  Rim, T. H. *et al.* Prediction of systemic biomarkers from retinal photographs: development and validation of deep-learning algorithms. *The Lancet Digital Health* **2**, e526–e536 (2020).

6.  Tarlan, B. & Kiratli, H. Subconjunctival hemorrhage: risk factors and potential indicators. *Clin. Ophthalmol.* **7**, 1163–1170 (2013).

7.  Hreidarsson, A. B. Pupil size in insulin-dependent diabetes. Relationship to duration, metabolic control, and long-term manifestations. *Diabetes* **31**, 442–448 (1982).

8.  Smith, S. E., Smith, S. A., Brown, P. M., Fox, C. & Sonksen, P. H. Pupillary signs in diabetic autonomic neuropathy. *BMJ* vol. 2 924–927 (1978).

9.  Banaee, T. *et al.* Distribution of Different Sized Ocular Surface Vessels in Diabetics and Normal Individuals. *J. Ophthalmic Vis. Res.* **12**, 361–367 (2017).

10. Iroshan, K. A. *et al.* Detection of Diabetes by Macrovascular Tortuosity of Superior Bulbar Conjunctiva. *Conf. Proc. IEEE Eng. Med. Biol. Soc.* **2018**, 1–4 (2018).

11. NCQA Comprehensive Diabetes Care. https://www.ncqa.org/hedis/measures/comprehensive-diabetes-care/.

12. Zangemeister, W. H., Gronow, T. & Grzyska, U. Pupillary responses to single and sinusoidal light stimuli in diabetic patients. *Neurol. Int.* **1**, e19 (2009).

13. Hashemi, H. *et al.* White-to-white corneal diameter distribution in an adult population. *J Curr*

*Ophthalmol* **27**, 21–24 (2015).

14. Rüfer, F., Schröder, A. & Erb, C. White-to-white corneal diameter: normal values in healthy humans obtained with the Orbscan II topography system. *Cornea* **24**, 259–261 (2005).

15. Piyasena, M. M. P. N. *et al.* Systematic review on barriers and enablers for access to diabetic retinopathy screening services in different income settings. *PLoS One* **14**, e0198979 (2019).

16. Stevenson, M., Lloyd-Jones, M., Morgan, M. Y. & Wong, R. *Diagnostic venepuncture: systematic review of adverse events*. (NIHR Journals Library, 2012).

17. Zhang, X. *et al.* Access to health care and control of ABCs of diabetes. *Diabetes Care* **35**, 1566–1571 (2012).

18. Klonoff, D. C. & Schwartz, D. M. An economic analysis of interventions for diabetes. *Diabetes Care* **23**, 390–404 (2000).

19. Alexander, G. C. *et al.* Use and Content of Primary Care Office-Based vs Telemedicine Care Visits During the COVID-19 Pandemic in the US. *JAMA Netw Open* **3**, e2021476 (2020).

20. Jalil, M., Ferenczy, S. R. & Shields, C. L. iPhone 4s and iPhone 5s Imaging of the Eye. *Ocul Oncol Pathol* **3**, 49–55 (2017).

21. Ludwig, C. A. *et al.* Training time and quality of smartphone-based anterior segment screening in rural India. *Clin. Ophthalmol.* **11**, 1301–1307 (2017).

22. Avram, R. *et al.* A digital biomarker of diabetes from smartphone-based vascular signals. *Nat. Med.* **26**, 1576–1582 (2020).

23. Worthen, D. M., Fenton, B. M., Rosen, P. & Zweifach, B. Morphometry of diabetic conjunctival blood vessels. *Ophthalmology* **88**, 655–657 (1981).

24. Danilova, A. I. [Blood circulation in the conjunctival blood vessels of patients with diabetes mellitus]. *Probl. Endokrinol.* **26**, 9–14 (1980).

25. Fenton, B. M., Zweifach, B. W. & Worthen, D. M. Quantitative morphometry of conjunctival microcirculation in diabetes mellitus. *Microvasc. Res.* **18**, 153–166 (1979).

26. Owen, C. G. *et al.* Diabetes and the tortuosity of vessels of the bulbar conjunctiva. *Ophthalmology* **115**, e27–32 (2008).

27. Owen, C. G., Newsom, R. S. B., Rudnicka, A. R., Ellis, T. J. & Woodward, E. G. Vascular response of the bulbar conjunctiva to diabetes and elevated blood pressure. *Ophthalmology* **112**, 1801–1808 (2005).

28. Khan, M. A. *et al.* A clinical correlation of conjunctival microangiopathy with grades of retinopathy in type 2 diabetes mellitus. *Armed Forces Med. J. India* **73**, 261–266 (2017).

29. Sharma, R., Gurunadh, V. S., Mathur, V. & Sati, A. Assessment of Conjunctival Vessel Calibre in Type-2 Diabetes Mellitus Patients. *International Journal of Enhanced Research in Medicines & Dental Care (IJERMDC)* **4**, (2017).

30. Santos, M. & Hofmann, R. J. Ocular Manifestations of Obstructive Sleep Apnea. *J. Clin. Sleep Med.* **13**, 1345–1348 (2017).

31. Cristescu Teodor, R. & Mihaltan, F. D. Eyelid laxity and sleep apnea syndrome: a review. *Rom J Ophthalmol* **63**, 2–9 (2019).

32. Scott, I. U. & Siatkowski, M. R. Thyroid eye disease. *Semin. Ophthalmol.* **14**, 52–61 (1999).

33. Dutton, J. J. Anatomic Considerations in Thyroid Eye Disease. *Ophthal. Plast. Reconstr. Surg.* **34**, S7–S12 (2018).

34. Chang, H.-C., Sung, C.-W. & Lin, M.-H. Serum lipids and risk of atherosclerosis in xanthelasma palpebrarum: A systematic review and meta-analysis. *J. Am. Acad. Dermatol.* **82**, 596–605 (2020).

35. Christoffersen, M. *et al.* Xanthelasmata, arcus corneae, and ischaemic vascular disease and death in general population: prospective cohort study. *BMJ* **343**, d5497 (2011).

36. To, W. J. *et al.* Real-time studies of hypertension using non-mydriatic fundus photography and computer-assisted intravital microscopy. *Clin. Hemorheol. Microcirc.* **53**, 267–279 (2013).

37. Mullaem, G. & Rosner, M. H. Ocular problems in the patient with end-stage renal disease. *Semin. Dial.* **25**, 403–407 (2012).

38. Klaassen-Broekema, N. & van Bijsterveld, O. P. Limbal and corneal calcification in patients with chronic renal failure. *Br. J. Ophthalmol.* **77**, 569–571 (1993).

39. Sharon, Y. & Schlesinger, N. Beyond Joints: a Review of Ocular Abnormalities in Gout and Hyperuricemia. *Curr. Rheumatol. Rep.* **18**, 37 (2016).

40. Wagner, S. K. *et al.* Insights into Systemic Disease through Retinal Imaging-Based Oculomics. *Transl. Vis. Sci. Technol.* **9**, 6 (2020).

41. Panwar, N. *et al.* Fundus Photography in the 21st Century--A Review of Recent Technological Advances and Their Implications for Worldwide Healthcare. *Telemed. J. E. Health.* **22**, 198–208 (2016).

42. Zhou, Y., Zhang, Y., Shi, K. & Wang, C. Body mass index and risk of diabetic retinopathy: A meta-analysis and systematic review. *Medicine* **96**, e6754 (2017).

43. Cuadros, J. & Bresnick, G. EyePACS: an adaptable telemedicine system for diabetic retinopathy screening. *J. Diabetes Sci. Technol.* **3**, 509–516 (2009).

44. Photographer Manual. https://www.eyepacs.org/photographer/protocol.jsp#external_photos.

45. Grading diabetic retinopathy from stereoscopic color fundus photographs--an extension of the modified Airlie House classification. ETDRS report number 10. Early Treatment Diabetic Retinopathy Study Research Group. *Ophthalmology* **98**, 786–806 (1991).

46. EyePACS digital retinal grading protocol. https://www.eyepacs.org/consultant/Clinical/grading/EyePACS-DIGITAL-RETINAL-IMAGE-GRADING.pdf.

47. Bora, A. *et al.* Predicting Risk of Developing Diabetic Retinopathy using Deep Learning. (2020).

48. Szegedy, C., Vanhoucke, V., Ioffe, S., Shlens, J. & Wojna, Z. Rethinking the Inception Architecture for Computer Vision. (2015).

49. Gulshan, V. *et al.* Development and Validation of a Deep Learning Algorithm for Detection of Diabetic Retinopathy in Retinal Fundus Photographs. *JAMA* **316**, 2402–2410 (2016).

50. DeLong, E. R., DeLong, D. M. & Clarke-Pearson, D. L. Comparing the areas under two or more correlated receiver operating characteristic curves: a nonparametric approach. *Biometrics* **44**, 837–845 (1988).

51. Selvaraju, R. R. *et al.* Grad-CAM: Visual Explanations from Deep Networks via Gradient-Based Localization. *2017 IEEE International Conference on Computer Vision (ICCV)* (2017) doi:10.1109/iccv.2017.74.

52. Springenberg, J. T., Dosovitskiy, A., Brox, T. & Riedmiller, M. Striving for Simplicity: The All Convolutional Net. *arXiv [cs.LG]* (2014).

53. Sundararajan, M., Taly, A. & Yan, Q. Axiomatic Attribution for Deep Networks. *arXiv [cs.LG]* (2017).

54. PAIR-code. PAIR-code/saliency. https://github.com/PAIR-code/saliency.

55. Chasan, J. E., Delaune, B., Maa, A. Y. & Lynch, M. G. Effect of a teleretinal screening program on eye care use and resources. *JAMA Ophthalmol.* **132**, 1045–1051 (2014).

# Supplementary Information

## Pupil / iris pupil segmentation model

A deep learning model was trained for pupil / iris pupil segmentation because several existing pupil/iris detection models did not perform well with images in our dataset upon testing. The iris and pupils were each modeled as axis-aligned ellipses with 4 parameters each: x / y center coordinates and x / y dimensions. All 8 parameters were represented as floating point numbers in the range [0,1], representing the fraction of the image width or height. A deep learning model was trained using a L2 regression loss to take as input the external eye image and output these 8 parameters. The architecture was otherwise identical to the disease prediction models. This model's hyperparameters are detailed in Supplementary Table 10, and results are presented in Supplementary Figure 4.

Supplementary Figures

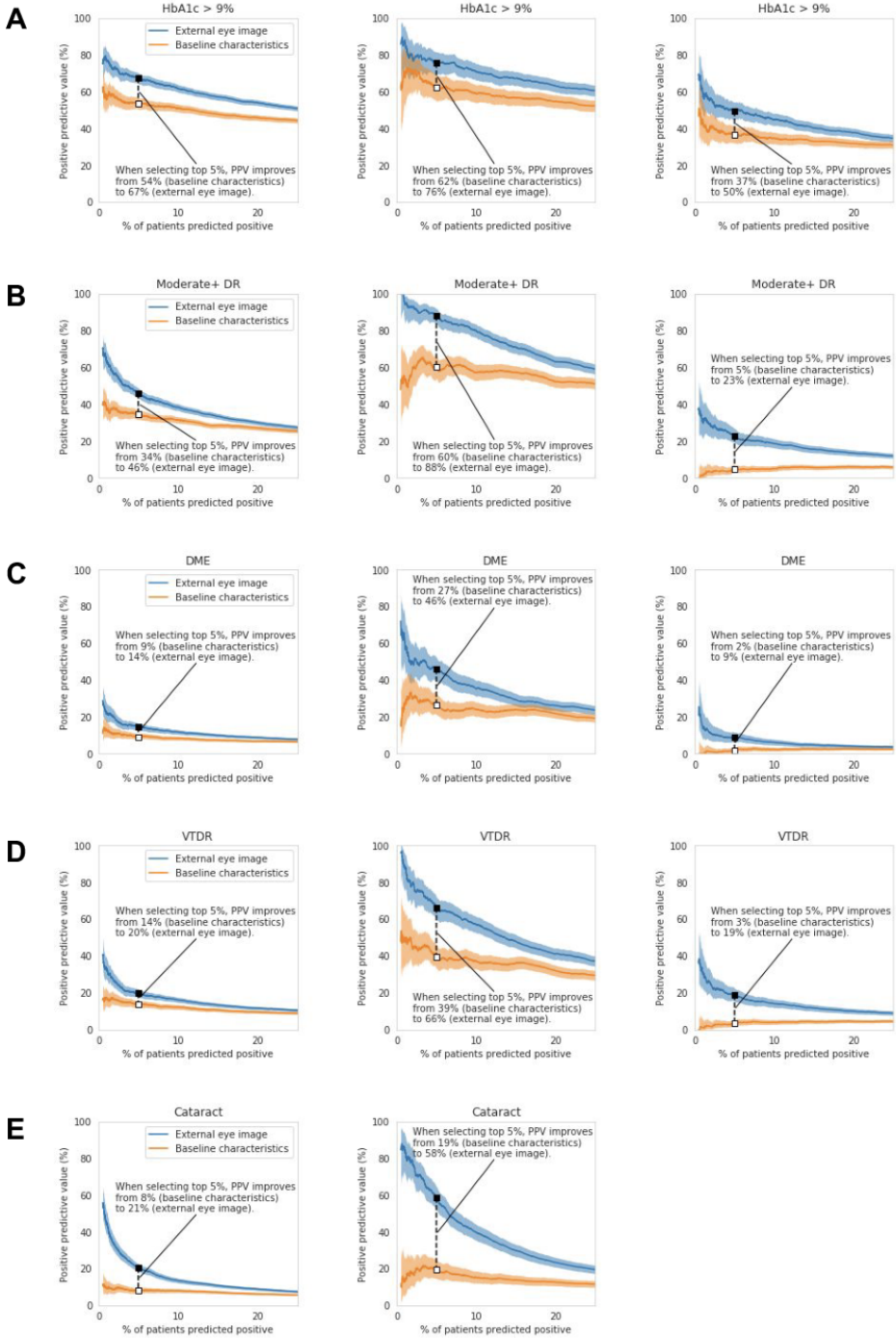

**Supplementary Figure 1. Curves of positive predictive value (PPV) as a function of threshold for various predictions using external eye images: (A) poor sugar control (HbA1c > 9%), (B) moderate-or-worse diabetic retinopathy (DR), (C) diabetic macular edema (DME), (D) vision-threatening DR (VTDR), and (E) a positive control: cataract.** In these plots, the x-axis indicates the percentage of patients predicted to be positive; for example 5% means the top 5% based on predicted likelihood was categorized to be "positive", and the respective curves indicate the PPV for that threshold. The curves are truncated at the extreme end (when only 0.5% of patients are predicted positive, confidence intervals are wide) to reduce noise and improve clarity.

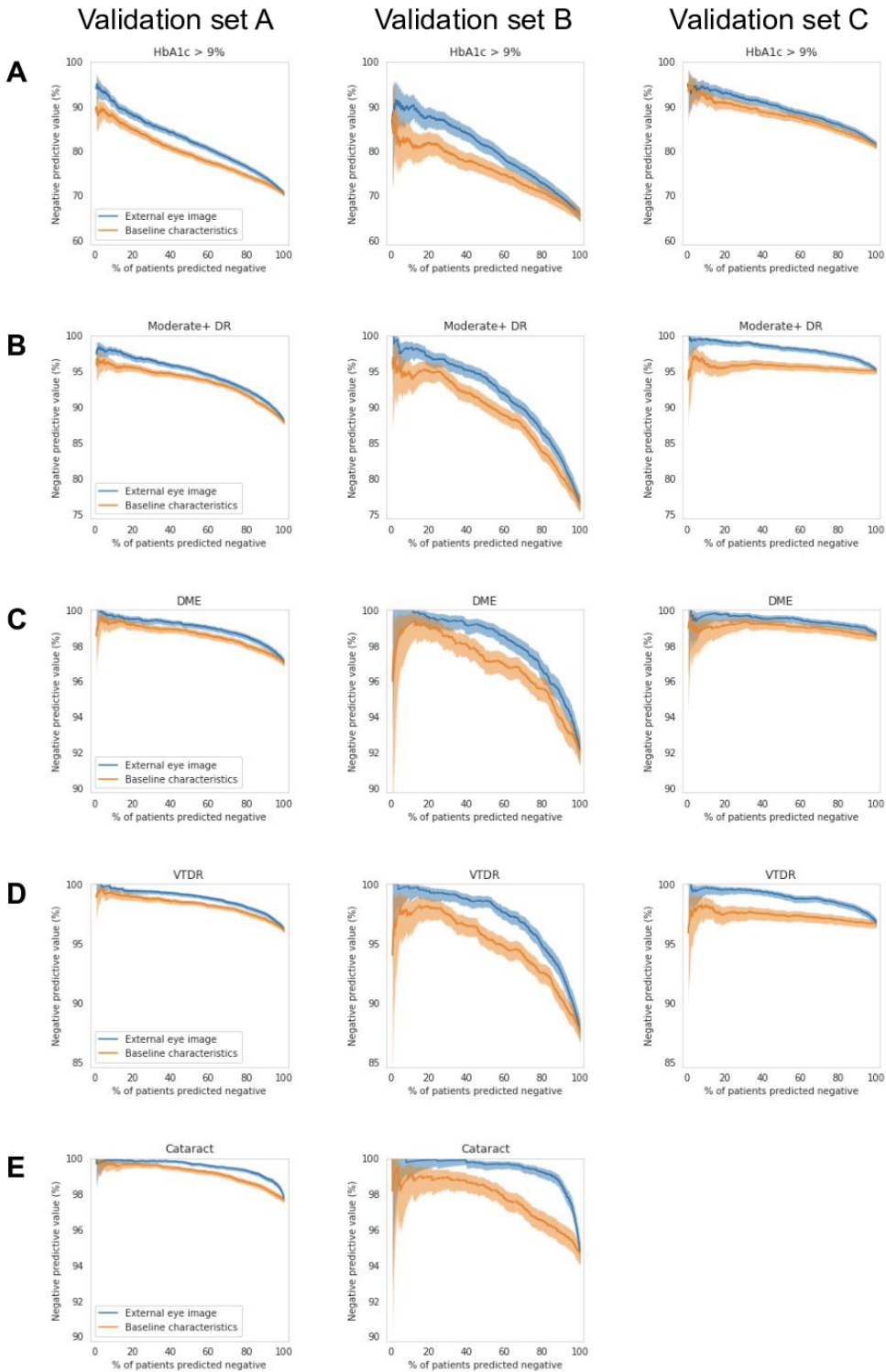

**Supplementary Figure 2. Curves of negative predictive values (NPV) as a function of threshold for various predictions using external eye images: (A) poor sugar control (HbA1c > 9%), (B) moderate-or-worse diabetic retinopathy (DR), (C) diabetic macular edema (DME), (D) vision-threatening DR (VTDR), and (E) a positive control: cataract.** This is the NPV equivalent of Supplementary Figure 1. The curves are truncated at the extreme end (when only 1% of patients are predicted negative, confidence intervals are wide) to reduce noise and improve clarity.

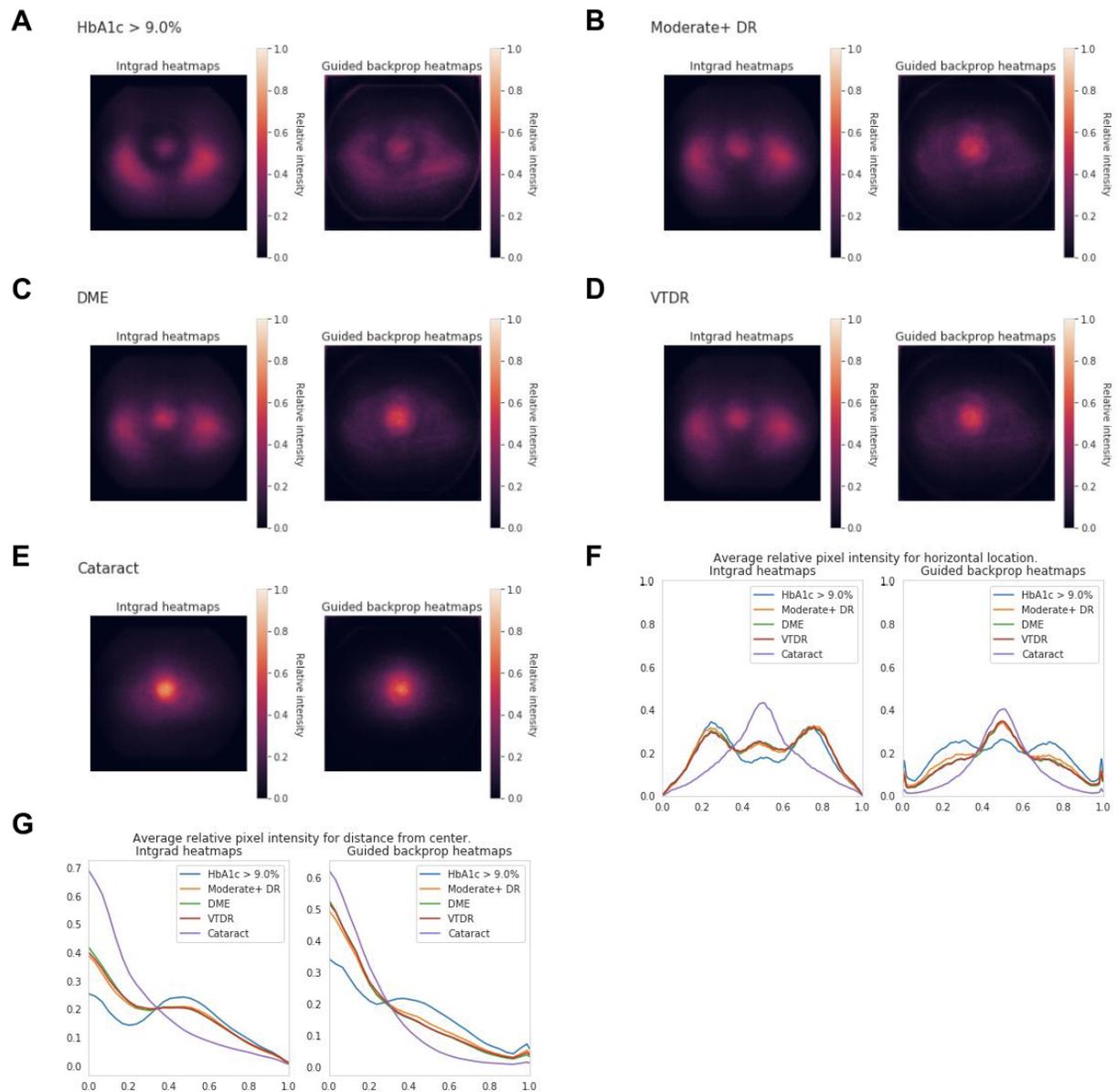

**Supplementary Figure 3. Saliency analysis illustrating the influence of various regions of the image towards the prediction.** Figures are generated in the same manner as in Figure 4, but with different saliency methods: Integrated Gradients in the middle, and guided backpropagation on the right.

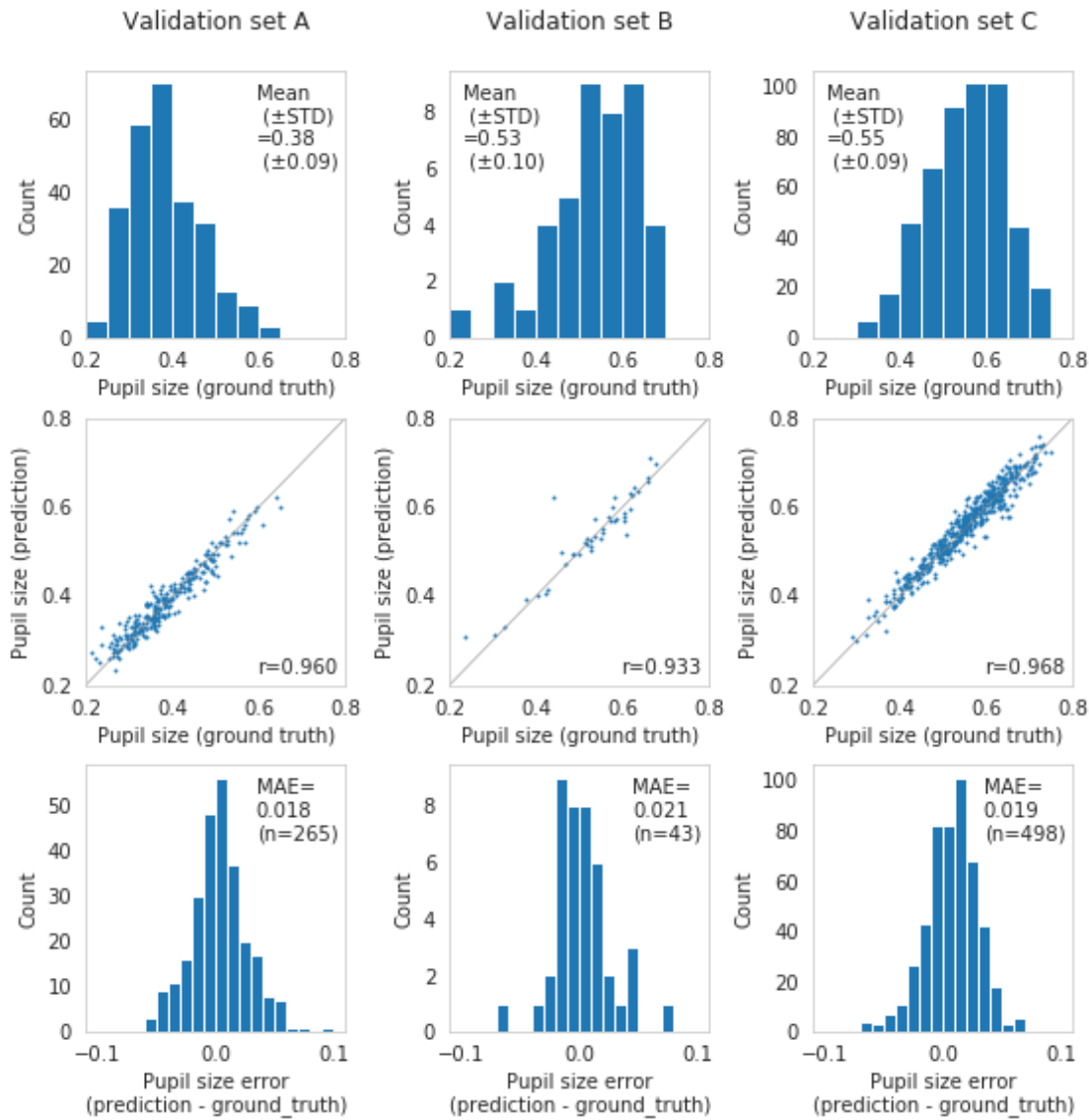

**Supplementary Figure 4. Accuracy of pupil size predictions by pupil/iris segmentation model.** Pupil size is represented as a fraction: the pupil radius divided by the iris radius. (Top row) Distribution of ground truth pupil size in each dataset. Note that validation set A contained non-dilated pupils, whereas pupils in validation sets B and C were dilated. (Middle row) Joint distribution of ground truth and predicted pupil size. Text indicates Pearson's correlation coefficient (r). (Bottom row) Distribution of errors (difference between predicted and ground truth pupil size). Abbreviations: STD = standard deviation, MAE = mean absolute error.

A

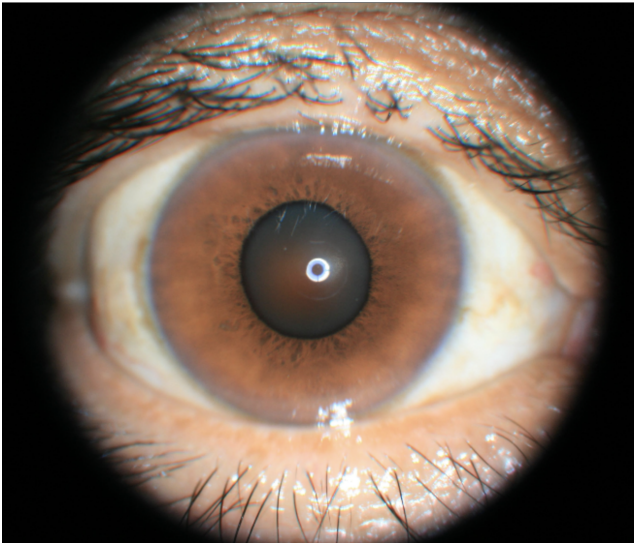

B

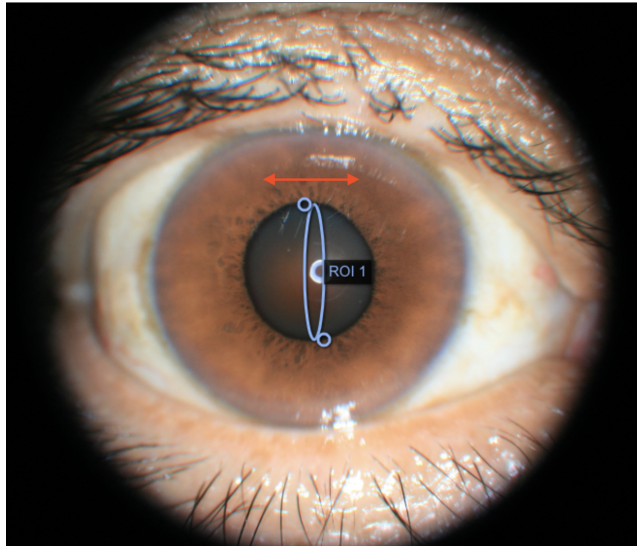

C

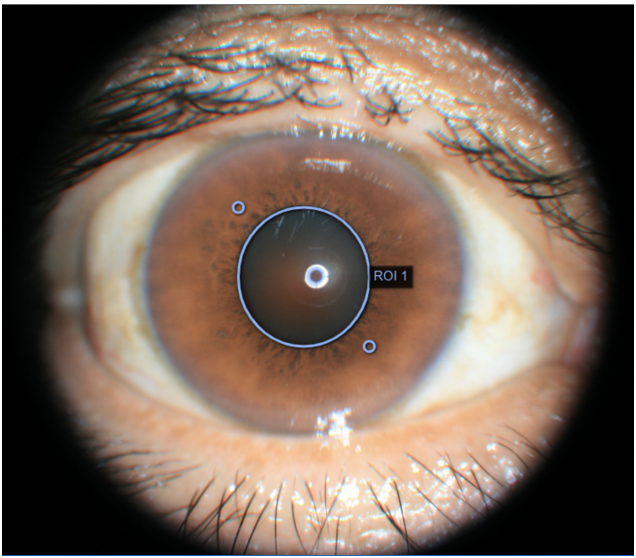

D

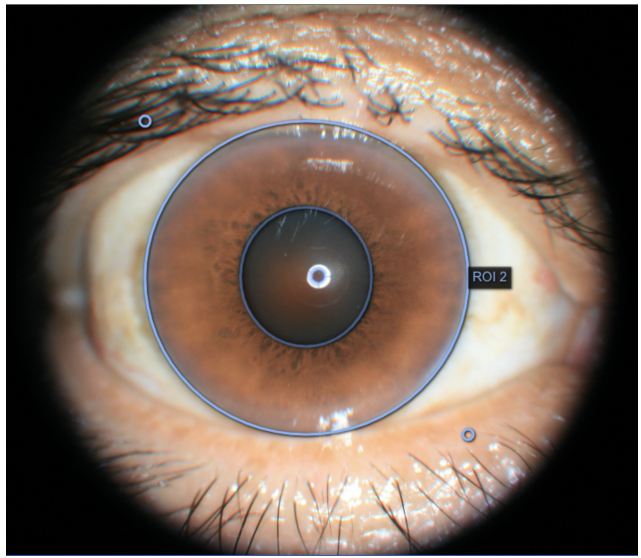

**Supplementary Figure 5. Pupil / iris segmentation tool involved drawing ellipses using a custom click and drag interface (A-C), to produce two ellipses: one around the pupil and another around the iris (D).**

# Supplementary Tables

**Supplementary Table 1. Area under receiver operating characteristics curve (AUC) of our models for the three validation sets.**

A

| Target | Total | Positive (%) | Baseline characteristics | DLS (external eye image) | Improvement | P-value for improvement |
|---|---|---|---|---|---|---|
| HbA1c > 7% | 21183 | 13544 (63.9%) | 65.0 (64.3-65.8) | 66.8 (66.1-67.6) | 1.8 (0.9-2.7) | < 0.001 |
| HbA1c > 8% | 21183 | 9347 (44.1%) | 65.2 (64.5-66.0) | 69.1 (68.4-69.9) | 3.9 (3.1-4.7) | < 0.001 |
| HbA1c > 9% | 21183 | 6268 (29.6%) | 64.8 (64.0-65.6) | 70.2 (69.4-70.9) | 5.4 (4.6-6.1) | < 0.001 |
| Mild+ DR | 26950 | 5144 (19.1%) | 68.9 (68.1-69.7) | 70.4 (69.6-71.2) | 1.4 (0.4-2.5) | 0.003 |
| Moderate+ DR | 26950 | 3247 (12.0%) | 71.0 (70.0-72.0) | 75.3 (74.4-76.2) | 4.3 (3.1-5.5) | < 0.001 |
| Severe+ DR | 26950 | 404 (1.5%) | 74.5 (72.0-76.9) | 83.2 (81.3-85.0) | 8.7 (5.7-11.7) | < 0.001 |
| DME | 26950 | 797 (3.0%) | 71.2 (69.4-73.0) | 78.0 (76.4-79.6) | 6.8 (4.7-9.0) | < 0.001 |
| VTDR | 26950 | 1042 (3.9%) | 72.7 (71.1-74.2) | 79.4 (78.1-80.8) | 6.8 (4.9-8.7) | < 0.001 |
| Cataract | 27415 | 639 (2.3%) | 76.5 (74.7-78.2) | 86.7 (85.3-88.1) | 10.2 (8.6-11.9) | < 0.001 |

B

| Target | Total | Positive (%) | Baseline characteristics | DLS (external eye image) | Improvement | P-value for improvement |
|---|---|---|---|---|---|---|
| HbA1c > 7% | 4120 | 2819 (68.4%) | 69.6 (67.9-71.3) | 74.4 (72.9-76.0) | 4.9 (3.0-6.7) | < 0.001 |
| HbA1c > 8% | 4120 | 1991 (48.3%) | 68.4 (66.8-70.0) | 74.0 (72.5-75.5) | 5.6 (4.0-7.2) | < 0.001 |
| HbA1c > 9% | 4120 | 1418 (34.4%) | 66.2 (64.4-67.9) | 73.4 (71.8-75.0) | 7.2 (5.6-8.9) | < 0.001 |
| Mild+ DR | 4982 | 1509 (30.3%) | 75.0 (73.6-76.5) | 80.3 (78.9-81.7) | 5.3 (3.6-6.9) | < 0.001 |
| Moderate+ DR | 4982 | 1172 (23.5%) | 77.2 (75.7-78.8) | 84.2 (82.9-85.5) | 6.9 (5.2-8.7) | < 0.001 |
| Severe+ DR | 4982 | 455 (9.1%) | 76.3 (74.1-78.5) | 89.0 (87.6-90.4) | 12.7 (10.3-15.1) | < 0.001 |
| DME | 4982 | 394 (7.9%) | 76.8 (74.5-79.1) | 85.2 (83.4-86.9) | 8.3 (5.6-11.1) | < 0.001 |
| VTDR | 4982 | 617 (12.4%) | 76.9 (75.0-78.8) | 87.1 (85.7-88.5) | 10.2 (8.0-12.4) | < 0.001 |
| Cataract | 5058 | 268 (5.3%) | 74.3 (71.5-77.2) | 93.4 (91.9-94.9) | 19.1 (16.2-22.0) | < 0.001 |

C

| Target | Total | Positive (%) | Baseline characteristics | DLS (external eye image) | Improvement | P-value for improvement |
|---|---|---|---|---|---|---|
| HbA1c > 7% | 8988 | 5127 (57.0%) | 57.8 (56.6-59.0) | 63.5 (62.4-64.7) | 5.7 (4.3-7.1) | < 0.001 |
| HbA1c > 8% | 8988 | 2962 (33.0%) | 61.7 (60.5-63.0) | 66.2 (65.0-67.4) | 4.5 (3.2-5.8) | < 0.001 |
| HbA1c > 9% | 8988 | 1685 (18.7%) | 65.3 (63.9-66.8) | 69.8 (68.4-71.2) | 4.4 (3.0-5.9) | < 0.001 |
| Mild+ DR | 9500 | 1006 (10.6%) | 52.8 (50.9-54.7) | 71.3 (69.6-73.1) | 18.6 (16.0-21.1) | < 0.001 |
| Moderate+ DR | 9500 | 462 (4.9%) | 54.1 (51.4-56.8) | 76.7 (74.5-78.9) | 22.6 (19.1-26.1) | < 0.001 |
| Severe+ DR | 9500 | 214 (2.3%) | 54.8 (51.1-58.5) | 79.9 (77.0-82.7) | 25.1 (20.1-30.0) | < 0.001 |
| DME | 9518 | 138 (1.4%) | 62.8 (58.2-67.5) | 75.8 (71.6-80.0) | 13.0 (6.8-19.1) | < 0.001 |
| VTDR | 9518 | 312 (3.3%) | 56.7 (53.5-59.9) | 79.0 (76.4-81.5) | 22.3 (18.1-26.5) | < 0.001 |

*Abbreviations: DR = diabetic retinopathy; DME = diabetic macular edema; HbA1c = hemoglobin A1c (glycated hemoglobin).

**Supplementary Table 2. Adjusted analysis for (A) validation set A, (B) validation set B, and (C) validation set C.** Bold indicates p-values < 0.05.

**A**

| Prediction target | No. of cases | | Odds ratio (95% CI), p-value | | | | | | |
|---|---|---|---|---|---|---|---|---|---|
| | Total | Positive (%) | Age (per decade) | Male | White* | Black* | Asian / Pacific islander* | Years with diabetes (per 5 years) | DLS (external eye image) |
| HbA1c > 7% | 21183 | 63.9 | **0.884 (0.859-0.910) p < 0.001** | **1.108 (1.044-1.177) p < 0.001** | 0.979 (0.907-1.056) p = 0.582 | 0.999 (0.913-1.093) p = 0.987 | 1.081 (0.980-1.193) p = 0.120 | **1.452 (1.409-1.497) p < 0.001** | **1.714 (1.654-1.775) p < 0.001** |
| HbA1c > 8% | 21183 | 44.1 | **0.907 (0.880-0.934) p < 0.001** | **1.070 (1.009-1.135) p = 0.025** | **0.926 (0.860-0.996) p = 0.040** | 0.956 (0.875-1.044) p = 0.316 | 1.029 (0.933-1.136) p = 0.564 | **1.326 (1.290-1.363) p < 0.001** | **1.943 (1.873-2.017) p < 0.001** |
| HbA1c > 9% | 21183 | 29.6 | **0.930 (0.900-0.961) p < 0.001** | **1.084 (1.017-1.155) p = 0.013** | **0.906 (0.836-0.981) p = 0.015** | 0.997 (0.906-1.097) p = 0.950 | 0.927 (0.831-1.033) p = 0.171 | **1.202 (1.168-1.238) p < 0.001** | **1.960 (1.884-2.039) p < 0.001** |
| Mild+ DR | 26950 | 19.1 | **0.906 (0.880-0.933) p < 0.001** | **1.157 (1.083-1.236) p < 0.001** | **0.758 (0.697-0.824) p < 0.001** | **1.100 (1.000-1.210) p = 0.049** | **1.309 (1.172-1.463) p < 0.001** | **1.638 (1.592-1.685) p < 0.001** | **1.798 (1.743-1.855) p < 0.001** |
| Moderate+ DR | 26950 | 12.0 | **0.910 (0.879-0.943) p < 0.001** | **1.197 (1.104-1.298) p < 0.001** | **0.771 (0.697-0.854) p < 0.001** | **0.882 (0.783-0.993) p = 0.038** | **1.149 (1.004-1.315) p = 0.044** | **1.673 (1.618-1.730) p < 0.001** | **1.892 (1.830-1.956) p < 0.001** |
| Severe+ DR | 26950 | 1.5 | **0.874 (0.802-0.953) p = 0.002** | **1.492 (1.214-1.834) p < 0.001** | 0.803 (0.613-1.052) p = 0.112 | 1.124 (0.824-1.533) p = 0.459 | **2.272 (1.694-3.047) p < 0.001** | **1.844 (1.698-2.003) p < 0.001** | **1.430 (1.371-1.491) p < 0.001** |
| DME | 26950 | 3.0 | **0.897 (0.842-0.956) p < 0.001** | **1.232 (1.063-1.427) p = 0.005** | **0.632 (0.519-0.771) p < 0.001** | 1.061 (0.867-1.297) p = 0.566 | 1.214 (0.954-1.545) p = 0.115 | **1.666 (1.571-1.767) p < 0.001** | **1.459 (1.402-1.519) p < 0.001** |
| VTDR | 26950 | 3.9 | **0.902 (0.853-0.955) p < 0.001** | **1.233 (1.081-1.405) p = 0.002** | **0.688 (0.578-0.819) p < 0.001** | 1.111 (0.924-1.335) p = 0.265 | **1.439 (1.169-1.772) p < 0.001** | **1.731 (1.642-1.825) p < 0.001** | **1.563 (1.507-1.621) p < 0.001** |
| Cataract | 27415 | 2.3 | **1.680 (1.545-1.826) p < 0.001** | **0.834 (0.700-0.994) p = 0.043** | **1.576 (1.241-2.002) p < 0.001** | **1.394 (1.047-1.855) p = 0.023** | **3.740 (2.943-4.752) p < 0.001** | **1.093 (1.018-1.173) p = 0.014** | **1.627 (1.565-1.691) p < 0.001** |

* Reference category: hispanic (the most prevalent race/ethnicity in this cohort)

**B**

| Prediction target | No. of cases | | Odds ratio (95% CI), p-value | | | | | | |
|---|---|---|---|---|---|---|---|---|---|
| | Total | Positive (%) | Age (per decade) | Male | White* | Black* | Asian / Pacific islander* | Years with diabetes (per 5 years) | DLS (external eye image) |
| HbA1c > 7% | 4120 | 68.4 | **0.869 (0.806-0.937) p < 0.001** | **1.178 (1.012-1.371) p = 0.034** | 0.886 (0.661-1.187) p = 0.416 | 0.848 (0.652-1.103) p = 0.218 | 1.621 (0.864-3.043) p = 0.133 | **1.422 (1.323-1.530) p < 0.001** | **2.242 (2.051-2.450) p < 0.001** |
| HbA1c > 8% | 4120 | 48.3 | **0.914 (0.850-0.984) p = 0.017** | **1.216 (1.056-1.400) p = 0.006** | 0.901 (0.677-1.199) p = 0.476 | **0.709 (0.553-0.909) p = 0.007** | 1.539 (0.848-2.792) p = 0.156 | **1.313 (1.233-1.398) p < 0.001** | **2.325 (2.130-2.537) p < 0.001** |
| HbA1c > 9% | 4120 | 34.4 | 0.952 (0.881-1.029) p = 0.212 | **1.160 (1.003-1.341) p = 0.045** | 0.912 (0.672-1.239) p = 0.556 | 0.839 (0.648-1.086) p = 0.182 | 1.535 (0.817-2.882) p = 0.183 | **1.178 (1.107-1.255) p < 0.001** | **2.254 (2.064-2.462) p < 0.001** |
| Mild+ DR | 4982 | 30.3 | 0.953 (0.888-1.023) p = 0.185 | **1.361 (1.172-1.582) p < 0.001** | 0.763 (0.573-1.016) p = 0.064 | 0.784 (0.603-1.021) p = 0.071 | **1.919 (1.038-3.550) p = 0.038** | **1.698 (1.593-1.810) p < 0.001** | **2.937 (2.702-3.193) p < 0.001** |
| Moderate+ DR | 4982 | 23.5 | **0.917 (0.848-0.992) p = 0.032** | **1.457 (1.233-1.722) p < 0.001** | **0.640 (0.457-0.897) p = 0.010** | **0.744 (0.555-0.998) p = 0.048** | 1.628 (0.820-3.232) p = 0.163 | **1.755 (1.638-1.881) p < 0.001** | **3.137 (2.875-3.423) p < 0.001** |
| Severe+ DR | 4982 | 9.1 | **0.862 (0.770-0.964) p = 0.010** | **1.439 (1.141-1.815) p = 0.002** | **0.476 (0.271-0.836) p = 0.010** | **0.545 (0.329-0.905) p = 0.019** | 0.169 (0.022-1.276) p = 0.085 | **1.615 (1.471-1.774) p < 0.001** | **2.515 (2.295-2.757) p < 0.001** |
| DME | 4982 | 7.9 | 1.038 (0.925-1.165) p = 0.522 | **1.646 (1.303-2.079) p < 0.001** | **0.320 (0.169-0.605) p < 0.001** | **0.557 (0.348-0.890) p = 0.014** | 0.626 (0.189-2.075) p = 0.443 | **1.734 (1.579-1.904) p < 0.001** | **1.958 (1.800-2.128) p < 0.001** |
| VTDR | 4982 | 12.4 | 0.930 (0.842-1.027) p = 0.151 | **1.555 (1.265-1.911) p < 0.001** | **0.427 (0.261-0.699) p < 0.001** | **0.590 (0.395-0.881) p = 0.010** | 0.581 (0.183-1.841) p = 0.356 | **1.715 (1.579-1.862) p < 0.001** | **2.587 (2.369-2.825) p < 0.001** |
| Cataract | 5058 | 5.3 | **1.494 (1.281-1.742) p < 0.001** | 1.290 (0.937-1.774) p = 0.118 | 1.325 (0.763-2.301) p = 0.318 | 0.547 (0.288-1.040) p = 0.066 | **3.167 (1.410-7.112) p = 0.005** | **1.262 (1.119-1.423) p < 0.001** | **2.775 (2.511-3.067) p < 0.001** |

* Reference category: hispanic (the most prevalent race/ethnicity in this cohort)

**C**

| Prediction target | No. of cases | | Odds ratio (95% CI), p-value | | |
|---|---|---|---|---|---|
| | Total | Positive (%) | Age (per decade) | Male | DLS (external eye image) |
| HbA1c > 7% | 8988 | 57.0 | **0.864 (0.825-0.906) p < 0.001** | 1.204 (0.976-1.486) p = 0.083 | **1.576 (1.503-1.652) p < 0.001** |
| HbA1c > 8% | 8988 | 33.0 | **0.828 (0.786-0.872) p < 0.001** | 1.188 (0.955-1.479) p = 0.122 | **1.701 (1.614-1.793) p < 0.001** |
| HbA1c > 9% | 8988 | 18.7 | **0.802 (0.751-0.856) p < 0.001** | 1.005 (0.785-1.286) p = 0.971 | **1.802 (1.694-1.917) p < 0.001** |
| Mild+ DR | 9500 | 10.6 | 0.942 (0.879-1.010) p = 0.093 | 1.420 (0.970-2.079) p = 0.072 | **1.923 (1.821-2.031) p < 0.001** |
| Moderate+ DR | 9500 | 4.9 | **0.873 (0.790-0.966) p = 0.008** | 1.642 (0.916-2.941) p = 0.096 | **1.968 (1.845-2.099) p < 0.001** |
| Severe+ DR | 9500 | 2.3 | **0.809 (0.704-0.929) p = 0.003** | **2.786 (1.015-7.644) p = 0.047** | **1.617 (1.502-1.740) p < 0.001** |
| DME | 9518 | 1.4 | **0.651 (0.549-0.772) p < 0.001** | 2.554 (0.833-7.837) p = 0.101 | **1.710 (1.546-1.891) p < 0.001** |
| VTDR | 9518 | 3.3 | **0.762 (0.676-0.857) p < 0.001** | **2.278 (1.056-4.913) p = 0.036** | **1.837 (1.712-1.970) p < 0.001** |

**Supplementary Table 3. Performance of adding predicted pupil size to baseline characteristics models for (A) validation set A, (B) validation set B, and (C) validation set C.** Note that per the other analyses, validation set C had these input variables fit directly on the validation set, and so over-estimates the AUC.

**A**

| Target | Total | Positive (%) | Baseline characteristics | Baseline characteristics + pupil size | DLS (external eye image) | Improvement | P-value for improvement |
|---|---|---|---|---|---|---|---|
| HbA1c > 9% | 21183 | 6268 (29.6%) | 64.8 (64.0-65.6) | 65.1 (64.3-65.9) | 70.2 (69.4-70.9) | 5.1 (4.3-5.9) | < 0.001 |
| Moderate+ DR | 26950 | 3247 (12.0%) | 71.0 (70.0-72.0) | 71.5 (70.6-72.5) | 75.3 (74.4-76.2) | 3.8 (2.6-4.9) | < 0.001 |
| DME | 26950 | 797 (3.0%) | 71.2 (69.4-73.0) | 71.7 (70.0-73.4) | 78.0 (76.4-79.6) | 6.3 (4.3-8.4) | < 0.001 |
| VTDR | 26950 | 1042 (3.9%) | 72.7 (71.1-74.2) | 73.4 (71.9-74.9) | 79.4 (78.1-80.8) | 6.0 (4.2-7.8) | < 0.001 |
| Cataract | 27415 | 639 (2.3%) | 76.5 (74.7-78.2) | 75.8 (74.1-77.6) | 86.7 (85.3-88.1) | 10.8 (9.1-12.6) | < 0.001 |

**B**

| Target | Total | Positive (%) | Baseline characteristics | Baseline characteristics + pupil size | DLS (external eye image) | Improvement | P-value for improvement |
|---|---|---|---|---|---|---|---|
| HbA1c > 9% | 4120 | 1418 (34.4%) | 66.2 (64.4-67.9) | 66.5 (64.8-68.3) | 73.4 (71.8-75.0) | 6.9 (5.3-8.4) | < 0.001 |
| Moderate+ DR | 4982 | 1172 (23.5%) | 77.2 (75.7-78.8) | 78.0 (76.6-79.5) | 84.2 (82.9-85.5) | 6.1 (4.5-7.8) | < 0.001 |
| DME | 4982 | 394 (7.9%) | 76.8 (74.5-79.1) | 78.0 (75.9-80.2) | 85.2 (83.4-86.9) | 7.1 (4.5-9.8) | < 0.001 |
| VTDR | 4982 | 617 (12.4%) | 76.9 (75.0-78.8) | 78.0 (76.2-79.9) | 87.1 (85.7-88.5) | 9.1 (7.0-11.1) | < 0.001 |
| Cataract | 5058 | 268 (5.3%) | 74.3 (71.5-77.2) | 75.9 (73.1-78.6) | 93.4 (91.9-94.9) | 17.6 (14.8-20.4) | < 0.001 |

**C**

| Target | Total | Positive (%) | Baseline characteristics | Baseline characteristics + pupil size | DLS (external eye image) | Improvement | P-value for improvement |
|---|---|---|---|---|---|---|---|
| HbA1c > 9% | 8988 | 1685 (18.7%) | 65.3 (63.9-66.8) | 65.8 (64.4-67.2) | 69.8 (68.4-71.2) | 4.0 (2.6-5.3) | < 0.001 |
| Moderate+ DR | 9500 | 462 (4.9%) | 54.1 (51.4-56.8) | 57.1 (54.6-59.7) | 76.7 (74.5-78.9) | 19.6 (16.5-22.6) | < 0.001 |
| DME | 9518 | 138 (1.4%) | 62.8 (58.2-67.5) | 63.4 (58.6-68.1) | 75.8 (71.6-80.0) | 12.4 (6.4-18.5) | < 0.001 |
| VTDR | 9518 | 312 (3.3%) | 56.7 (53.5-59.9) | 60.6 (57.4-63.7) | 79.0 (76.4-81.5) | 18.4 (14.6-22.2) | < 0.001 |

**Supplementary Table 4. Adjusted analysis including pupil size for (A) validation set A, (B) validation set B, and (C) validation set C.** Similar to supplementary Table 2.

**A**

| Prediction target | Total | Positive (%) | Age (per decade) | Years with diabetes (per 5 years) | Male | White | Black | Asian / Pacific islander | Pupil radius relative to iris | DLS (external eye image) |
|---|---|---|---|---|---|---|---|---|---|---|
| HbA1c > 7% | 21183 | 63.9 | 0.884 (0.858-0.910) p < 0.001 | 1.452 (1.409-1.497) p < 0.001 | 1.108 (1.043-1.177) p < 0.001 | 0.979 (0.907-1.056) p = 0.580 | 0.999 (0.912-1.093) p = 0.979 | 1.081 (0.980-1.193) p = 0.120 | 0.998 (0.956-1.042) p = 0.935 | 1.713 (1.653-1.776) p < 0.001 |
| HbA1c > 8% | 21183 | 44.1 | 0.902 (0.875-0.930) p < 0.001 | 1.325 (1.289-1.362) p < 0.001 | 1.066 (1.005-1.131) p = 0.034 | 0.924 (0.859-0.995) p = 0.037 | 0.949 (0.869-1.037) p = 0.250 | 1.028 (0.932-1.134) p = 0.582 | 0.971 (0.931-1.013) p = 0.175 | 1.935 (1.864-2.009) p < 0.001 |
| HbA1c > 9% | 21183 | 29.6 | 0.919 (0.889-0.951) p < 0.001 | 1.201 (1.167-1.236) p < 0.001 | 1.076 (1.009-1.147) p = 0.025 | 0.904 (0.834-0.979) p = 0.013 | 0.983 (0.893-1.082) p = 0.724 | 0.923 (0.827-1.029) p = 0.148 | 0.938 (0.896-0.981) p = 0.006 | 1.944 (1.868-2.023) p < 0.001 |
| Mild+ DR | 26950 | 19.1 | 0.896 (0.870-0.923) p < 0.001 | 1.633 (1.587-1.680) p < 0.001 | 1.141 (1.067-1.220) p < 0.001 | 0.760 (0.699-0.827) p < 0.001 | 1.076 (0.978-1.185) p = 0.132 | 1.309 (1.171-1.463) p < 0.001 | 0.901 (0.861-0.943) p < 0.001 | 1.793 (1.738-1.850) p < 0.001 |
| Moderate+ DR | 26950 | 12.0 | 0.895 (0.864-0.928) p < 0.001 | 1.668 (1.613-1.725) p < 0.001 | 1.172 (1.080-1.271) p < 0.001 | 0.775 (0.700-0.858) p < 0.001 | 0.854 (0.758-0.963) p = 0.010 | 1.146 (1.001-1.312) p = 0.049 | 0.851 (0.804-0.900) p < 0.001 | 1.879 (1.818-1.943) p < 0.001 |
| Severe+ DR | 26950 | 1.5 | 0.860 (0.788-0.939) p < 0.001 | 1.832 (1.687-1.990) p < 0.001 | 1.453 (1.182-1.788) p < 0.001 | 0.808 (0.617-1.059) p = 0.122 | 1.089 (0.797-1.486) p = 0.593 | 2.261 (1.686-3.033) p < 0.001 | 0.832 (0.721-0.961) p = 0.013 | 1.422 (1.363-1.483) p < 0.001 |
| DME | 26950 | 3.0 | 0.881 (0.826-0.940) p < 0.001 | 1.653 (1.559-1.753) p < 0.001 | 1.197 (1.033-1.388) p = 0.017 | 0.637 (0.522-0.777) p < 0.001 | 1.022 (0.835-1.252) p = 0.831 | 1.208 (0.949-1.537) p = 0.125 | 0.821 (0.740-0.910) p < 0.001 | 1.456 (1.399-1.516) p < 0.001 |
| VTDR | 26950 | 3.9 | 0.888 (0.839-0.940) p < 0.001 | 1.721 (1.633-1.815) p < 0.001 | 1.204 (1.055-1.374) p = 0.006 | 0.692 (0.581-0.823) p < 0.001 | 1.078 (0.896-1.297) p = 0.426 | 1.433 (1.164-1.765) p < 0.001 | 0.848 (0.773-0.930) p < 0.001 | 1.554 (1.498-1.613) p < 0.001 |
| Cataract | 27415 | 2.3 | 1.652 (1.519-1.798) p < 0.001 | 1.083 (1.009-1.162) p = 0.027 | 0.809 (0.678-0.966) p = 0.019 | 1.577 (1.242-2.004) p < 0.001 | 1.347 (1.011-1.794) p = 0.042 | 3.769 (2.966-4.791) p < 0.001 | 0.828 (0.730-0.939) p = 0.003 | 1.634 (1.572-1.700) p < 0.001 |

**B**

| Prediction target | Total | Positive (%) | Age (per decade) | Years with diabetes (per 5 years) | Male | White | Black | Asian / Pacific islander | Pupil radius relative to iris | DLS (external eye image) |
|---|---|---|---|---|---|---|---|---|---|---|
| HbA1c > 7% | 4120 | 68.4 | 0.865 (0.801-0.935) p < 0.001 | 1.420 (1.321-1.528) p < 0.001 | 1.175 (1.009-1.368) p = 0.039 | 0.879 (0.655-1.180) p = 0.391 | 0.841 (0.645-1.097) p = 0.201 | 1.607 (0.856-3.019) p = 0.140 | 0.978 (0.884-1.082) p = 0.664 | 2.236 (2.044-2.446) p < 0.001 |
| HbA1c > 8% | 4120 | 48.3 | 0.915 (0.848-0.987) p = 0.022 | 1.313 (1.233-1.399) p < 0.001 | 1.217 (1.056-1.402) p = 0.007 | 0.902 (0.677-1.202) p = 0.483 | 0.710 (0.553-0.911) p = 0.007 | 1.541 (0.849-2.797) p = 0.155 | 1.003 (0.913-1.103) p = 0.943 | 2.326 (2.129-2.542) p < 0.001 |
| HbA1c > 9% | 4120 | 34.4 | 0.949 (0.876-1.029) p = 0.204 | 1.178 (1.106-1.254) p < 0.001 | 1.158 (1.001-1.339) p = 0.048 | 0.909 (0.668-1.236) p = 0.542 | 0.836 (0.646-1.083) p = 0.176 | 1.529 (0.814-2.873) p = 0.187 | 0.987 (0.895-1.089) p = 0.799 | 2.249 (2.055-2.461) p < 0.001 |
| Mild+ DR | 4982 | 30.3 | 0.938 (0.873-1.008) p = 0.080 | 1.683 (1.579-1.795) p < 0.001 | 1.334 (1.147-1.552) p < 0.001 | 0.742 (0.557-0.989) p = 0.042 | 0.757 (0.581-0.986) p = 0.039 | 1.868 (1.013-3.446) p = 0.045 | 0.883 (0.802-0.972) p = 0.011 | 2.920 (2.686-3.175) p < 0.001 |
| Moderate+ DR | 4982 | 23.5 | 0.891 (0.822-0.965) p = 0.005 | 1.732 (1.616-1.857) p < 0.001 | 1.408 (1.190-1.666) p < 0.001 | 0.610 (0.435-0.855) p = 0.004 | 0.702 (0.524-0.942) p = 0.018 | 1.573 (0.798-3.100) p = 0.191 | 0.802 (0.721-0.892) p < 0.001 | 3.099 (2.840-3.382) p < 0.001 |
| Severe+ DR | 4982 | 9.1 | 0.841 (0.751-0.943) p = 0.003 | 1.599 (1.456-1.757) p < 0.001 | 1.402 (1.110-1.770) p = 0.005 | 0.461 (0.263-0.808) p = 0.007 | 0.521 (0.314-0.865) p = 0.012 | 0.176 (0.024-1.283) p = 0.087 | 0.836 (0.722-0.968) p = 0.017 | 2.478 (2.259-2.717) p < 0.001 |
| DME | 4982 | 7.9 | 0.997 (0.888-1.121) p = 0.966 | 1.698 (1.545-1.865) p < 0.001 | 1.566 (1.237-1.981) p < 0.001 | 0.306 (0.163-0.575) p < 0.001 | 0.512 (0.320-0.821) p = 0.005 | 0.624 (0.191-2.043) p = 0.436 | 0.734 (0.635-0.848) p < 0.001 | 1.946 (1.789-2.115) p < 0.001 |
| VTDR | 4982 | 12.4 | 0.894 (0.808-0.989) p = 0.029 | 1.685 (1.550-1.831) p < 0.001 | 1.484 (1.206-1.826) p < 0.001 | 0.401 (0.246-0.656) p < 0.001 | 0.548 (0.367-0.820) p = 0.003 | 0.579 (0.185-1.811) p = 0.348 | 0.746 (0.655-0.848) p < 0.001 | 2.543 (2.329-2.777) p < 0.001 |
| Cataract | 5058 | 5.3 | 1.442 (1.236-1.683) p < 0.001 | 1.220 (1.080-1.377) p = 0.001 | 1.199 (0.869-1.654) p = 0.269 | 1.221 (0.699-2.131) p = 0.483 | 0.482 (0.253-0.917) p = 0.026 | 2.895 (1.278-6.560) p = 0.011 | 0.674 (0.558-0.814) p < 0.001 | 2.738 (2.476-3.027) p < 0.001 |

**C**

| Prediction target | Total | Positive (%) | Age (per decade) | Male | Pupil radius relative to iris | DLS (external eye image) |
|---|---|---|---|---|---|---|
| HbA1c > 7% | 8988 | 57.0 | 0.873 (0.832-0.916) p < 0.001 | 1.200 (0.973-1.481) p = 0.089 | 1.613 (0.902-2.886) p = 0.107 | 1.594 (1.517-1.674) p < 0.001 |
| HbA1c > 8% | 8988 | 33.0 | 0.832 (0.788-0.878) p < 0.001 | 1.186 (0.953-1.476) p = 0.126 | 1.217 (0.653-2.266) p = 0.537 | 1.709 (1.618-1.805) p < 0.001 |
| HbA1c > 9% | 8988 | 18.7 | 0.805 (0.752-0.861) p < 0.001 | 1.003 (0.783-1.284) p = 0.983 | 1.176 (0.553-2.502) p = 0.673 | 1.809 (1.697-1.928) p < 0.001 |
| Mild+ DR | 9500 | 10.6 | 0.930 (0.867-0.997) p = 0.041 | 1.438 (0.981-2.107) p = 0.063 | 0.339 (0.142-0.806) p = 0.014 | 1.907 (1.805-2.015) p < 0.001 |
| Moderate+ DR | 9500 | 4.9 | 0.853 (0.771-0.943) p = 0.002 | 1.690 (0.942-3.031) p = 0.079 | 0.130 (0.038-0.444) p = 0.001 | 1.939 (1.817-2.069) p < 0.001 |
| Severe+ DR | 9500 | 2.3 | 0.772 (0.672-0.888) p < 0.001 | 2.894 (1.054-7.947) p = 0.039 | 0.021 (0.004-0.114) p < 0.001 | 1.555 (1.444-1.675) p < 0.001 |
| DME | 9518 | 1.4 | 0.646 (0.544-0.767) p < 0.001 | 2.576 (0.841-7.890) p = 0.097 | 0.455 (0.051-4.080) p = 0.482 | 1.699 (1.536-1.880) p < 0.001 |
| VTDR | 9518 | 3.3 | 0.743 (0.660-0.837) p < 0.001 | 2.351 (1.090-5.071) p = 0.029 | 0.105 (0.024-0.459) p = 0.003 | 1.794 (1.670-1.926) p < 0.001 |

**Supplementary Table 5. Subgroup analysis for HbA1c > 9% in validation set A.** Note that the sum of some subgroups exceed the total because some patients have multiple visits over time that can fall under different groups (e.g., age, years with diabetes, cataract etc).

| Category | Subgroup | No. of cases | | AUC (%) (95% CI) | | | |
|---|---|---|---|---|---|---|---|
| | | Total | Positive (%) | Baseline characteristics | DLS (external eye image) | Improvement | P-value for improvement |
| All | | 21183 | 6268 (29.6%) | 64.8 (64.0-65.6) | 70.2 (69.4-70.9) | 5.4 (4.6-6.1) | < 0.001 |
| Age | ≤50 years old | 8081 | 3093 (38.3%) | 60.9 (59.7-62.2) | 67.9 (66.7-69.1) | 7.0 (5.6-8.3) | < 0.001 |
| | 50-60 years old | 6989 | 2026 (29.0%) | 60.8 (59.3-62.2) | 67.0 (65.6-68.3) | 6.2 (4.5-7.8) | < 0.001 |
| | >60 years old | 6542 | 1271 (19.4%) | 61.6 (59.9-63.3) | 67.4 (65.8-69.1) | 5.8 (3.9-7.8) | < 0.001 |
| Sex | Male | 10019 | 3219 (32.1%) | 64.1 (63.0-65.3) | 69.3 (68.2-70.4) | 5.1 (4.0-6.2) | < 0.001 |
| | Female | 11154 | 3096 (27.8%) | 64.9 (63.8-66.1) | 70.4 (69.3-71.5) | 5.5 (4.3-6.6) | < 0.001 |
| Race / ethnicity | Hispanic | 8687 | 3007 (34.6%) | 63.4 (62.2-64.6) | 70.2 (69.1-71.3) | 6.8 (5.6-8.0) | < 0.001 |
| | White | 5559 | 1500 (27.0%) | 63.1 (61.5-64.8) | 66.1 (64.5-67.7) | 3.0 (1.4-4.5) | < 0.001 |
| | Black | 3147 | 879 (27.9%) | 65.7 (63.6-67.8) | 66.8 (64.7-68.9) | 1.0 (-1.2-3.3) | 0.178 |
| | Asian / Pacific islander | 2615 | 593 (22.7%) | 65.8 (63.3-68.2) | 76.0 (73.8-78.2) | 10.2 (7.8-12.6) | < 0.001 |
| No. years with diabetes | ≤5 | 12095 | 3115 (25.8%) | 63.6 (62.5-64.7) | 69.6 (68.5-70.7) | 6.0 (5.0-7.0) | < 0.001 |
| | 5-10 | 5557 | 1799 (32.4%) | 64.8 (63.3-66.3) | 71.5 (70.0-72.9) | 6.7 (5.3-8.0) | < 0.001 |
| | >10 | 4914 | 1778 (36.2%) | 63.6 (62.0-65.2) | 69.7 (68.2-71.2) | 6.1 (4.5-7.6) | < 0.001 |
| Cataract* | Present | 579 | 152 (26.3%) | 65.5 (60.4-70.6) | 69.5 (64.5-74.4) | 4.0 (-1.0-8.9) | 0.059 |
| | Absent | 20679 | 6175 (29.9%) | 64.7 (63.9-65.5) | 70.1 (69.3-70.9) | 5.4 (4.6-6.2) | < 0.001 |
| Pupil diameter* | Small pupil | 6186 | 1915 (31.0%) | 63.9 (62.4-65.3) | 70.1 (68.7-71.5) | 6.2 (4.7-7.8) | < 0.001 |
| | Medium pupil | 11152 | 3289 (29.5%) | 65.3 (64.2-66.4) | 69.8 (68.7-70.9) | 4.5 (3.5-5.6) | < 0.001 |
| | Large pupil | 5537 | 1558 (28.1%) | 65.9 (64.4-67.5) | 71.4 (70.0-72.9) | 5.5 (4.1-6.9) | < 0.001 |

*Because two eyes of the same patient can have different cataract statuses and pupil sizes, for this analysis we selected a random eye that fit this criteria per patient. In other words, in contrast to other analysis in this paper, these HbA1c predictions are from a single eye instead of being the average of two eyes.

**Supplementary Table 6. Subgroup analysis for moderate+ diabetic retinopathy (DR) in validation set A.** Formatting as per Supplementary Table 5. Bold indicates there is a >5% decrease in AUC for the DLS in that subgroup compared to the entire cohort.

| Category | Subgroup | No. of cases | | AUC (%) (95% CI) | | | |
|---|---|---|---|---|---|---|---|
| | | Total | Positive (%) | Baseline characteristics | DLS (external eye image) | Improvement | P-value for improvement |
| All | | 26950 | 3247 (12.0%) | 71.0 (70.0-72.0) | 75.3 (74.4-76.2) | 4.3 (3.1-5.5) | < 0.001 |
| Age | ≤50 years old | 10075 | 1198 (11.9%) | 74.1 (72.6-75.6) | 77.5 (76.1-78.9) | 3.3 (1.5-5.2) | < 0.001 |
| | 50-60 years old | 8937 | 1118 (12.5%) | 69.8 (68.1-71.5) | 73.7 (72.2-75.3) | 4.0 (1.9-6.0) | < 0.001 |
| | >60 years old | 8415 | 911 (10.8%) | 69.6 (67.8-71.4) | 74.1 (72.5-75.8) | 4.5 (2.2-6.8) | < 0.001 |
| Sex | Male | 12659 | 1650 (13.0%) | 70.2 (68.8-71.6) | 75.4 (74.2-76.6) | 5.2 (3.5-6.9) | < 0.001 |
| | Female | 14301 | 1520 (10.6%) | 71.1 (69.6-72.5) | 76.0 (74.7-77.3) | 4.9 (3.2-6.6) | < 0.001 |
| Race / ethnicity | Hispanic | 11177 | 1456 (13.0%) | 71.6 (70.1-73.0) | 77.3 (76.0-78.6) | 5.7 (4.0-7.5) | < 0.001 |
| | White | 7528 | 715 (9.5%) | 72.2 (70.1-74.2) | 74.6 (72.7-76.5) | 2.4 (-0.1-4.9) | 0.028 |
| | **Black** | **4094** | **470 (11.5%)** | **68.2 (65.7-70.8)** | **69.1 (66.5-71.7)** | **0.8 (-2.4-4.1)** | **0.303** |
| | Asian / Pacific islander | 2882 | 354 (12.3%) | 65.5 (62.3-68.6) | 77.7 (75.2-80.3) | 12.3 (8.6-15.9) | < 0.001 |
| No. years with diabetes | ≤5 | 15059 | 922 (6.1%) | 59.7 (57.8-61.6) | 75.1 (73.4-76.7) | 15.4 (13.1-17.6) | < 0.001 |
| | 5-10 | 7023 | 882 (12.6%) | 60.2 (58.2-62.2) | 73.7 (71.9-75.4) | 13.5 (11.0-15.9) | < 0.001 |
| | >10 | 6457 | 1565 (24.2%) | 60.0 (58.4-61.5) | 71.9 (70.5-73.4) | 12.0 (10.0-14.0) | < 0.001 |
| Cataract | Present | 656 | 71 (10.8%) | 66.4 (59.3-73.5) | 74.2 (68.9-79.5) | 7.8 (-0.2-15.8) | 0.028 |
| | Absent | 26413 | 3099 (11.7%) | 71.1 (70.2-72.1) | 75.8 (74.9-76.7) | 4.7 (3.5-5.9) | < 0.001 |
| HbA1c (%) | ≤7 | 8130 | 360 (4.4%) | 66.8 (63.7-69.8) | 72.0 (69.2-74.9) | 5.3 (1.8-8.8) | 0.002 |
| | 7-9 | 7972 | 990 (12.4%) | 68.7 (66.9-70.5) | 73.8 (72.1-75.4) | 5.0 (2.8-7.3) | < 0.001 |
| | >9 | 6665 | 1361 (20.4%) | 67.8 (66.2-69.4) | 72.2 (70.7-73.7) | 4.4 (2.4-6.4) | < 0.001 |
| Pupil diameter | Small pupil | 10645 | 1426 (13.4%) | 69.5 (68.1-71.0) | 73.4 (72.0-74.8) | 3.9 (2.1-5.7) | < 0.001 |
| | Medium pupil | 17565 | 2162 (12.3%) | 70.9 (69.7-72.0) | 75.5 (74.4-76.5) | 4.6 (3.2-6.0) | < 0.001 |
| | Large pupil | 9506 | 853 (9.0%) | 72.3 (70.4-74.2) | 75.7 (74.0-77.3) | 3.4 (1.1-5.6) | 0.001 |

**Supplementary Table 7. Subgroup analysis for diabetic macula edema (DME) in validation set A.**
Formatting as per Supplementary Table 5.

| Category | Subgroup | No. of cases | | AUC (%) (95% CI) | | | |
|---|---|---|---|---|---|---|---|
| | | Total | Positive (%) | Baseline characteristics | DLS (external eye image) | Improvement | P-value for improvement |
| All | | 26950 | 797 (3.0%) | 71.2 (69.4-73.0) | 78.0 (76.4-79.6) | 6.8 (4.7-9.0) | < 0.001 |
| Age | ≤50 years old | 10075 | 280 (2.8%) | 73.5 (70.5-76.5) | 77.9 (75.2-80.5) | 4.4 (0.9-7.9) | 0.007 |
| | 50-60 years old | 8937 | 301 (3.4%) | 69.5 (66.6-72.4) | 77.7 (75.0-80.4) | 8.2 (4.7-11.7) | < 0.001 |
| | >60 years old | 8415 | 230 (2.7%) | 68.3 (64.9-71.7) | 74.5 (71.1-77.9) | 6.2 (1.7-10.7) | 0.003 |
| Sex | Male | 12659 | 426 (3.4%) | 67.7 (65.1-70.4) | 76.9 (74.7-79.1) | 9.2 (6.0-12.3) | < 0.001 |
| | Female | 14301 | 404 (2.8%) | 72.6 (70.2-75.0) | 77.7 (75.4-80.0) | 5.1 (2.1-8.1) | < 0.001 |
| Race / ethnicity | Hispanic | 11177 | 361 (3.2%) | 70.5 (67.8-73.1) | 79.8 (77.5-82.0) | 9.3 (6.2-12.4) | < 0.001 |
| | White | 7528 | 153 (2.0%) | 72.2 (68.3-76.1) | 75.0 (71.1-78.9) | 2.8 (-2.1-7.7) | 0.130 |
| | Black | 4094 | 146 (3.6%) | 69.9 (65.5-74.2) | 74.2 (70.2-78.2) | 4.3 (-1.2-9.8) | 0.063 |
| | Asian / Pacific islander | 2882 | 93 (3.2%) | 69.2 (63.8-74.6) | 80.1 (75.3-84.9) | 10.9 (4.8-17.1) | < 0.001 |
| No. years with diabetes | ≤5 | 15059 | 219 (1.5%) | 61.5 (57.7-65.3) | 78.3 (75.3-81.4) | 16.8 (12.2-21.5) | < 0.001 |
| | 5-10 | 7023 | 204 (2.9%) | 56.9 (52.9-60.9) | 77.8 (74.8-80.7) | 20.9 (16.3-25.5) | < 0.001 |
| | >10 | 6457 | 412 (6.4%) | 58.3 (55.5-61.2) | 73.0 (70.5-75.4) | 14.6 (11.3-18.0) | < 0.001 |
| Cataract | Present | 656 | 14 (2.1%) | 61.6 (45.4-77.8) | 73.5 (58.7-88.2) | 11.9 (-7.6-31.4) | 0.116 |
| | Absent | 26413 | 798 (3.0%) | 71.5 (69.7-73.3) | 77.6 (76.1-79.2) | 6.1 (4.0-8.3) | < 0.001 |
| HbA1c (%) | ≤7 | 8130 | 62 (0.8%) | 68.2 (61.0-75.3) | 74.4 (68.3-80.5) | 6.2 (-1.6-14.1) | 0.061 |
| | 7-9 | 7972 | 233 (2.9%) | 68.1 (64.6-71.5) | 78.2 (75.2-81.2) | 10.1 (5.9-14.4) | < 0.001 |
| | >9 | 6665 | 340 (5.1%) | 66.8 (64.0-69.7) | 74.7 (72.1-77.3) | 7.9 (4.4-11.4) | < 0.001 |
| Pupil diameter | Small pupil | 10645 | 368 (3.5%) | 66.7 (64.1-69.4) | 77.0 (74.7-79.3) | 10.2 (6.9-13.6) | < 0.001 |
| | Medium pupil | 17565 | 540 (3.1%) | 69.9 (67.7-72.1) | 77.0 (75.0-78.9) | 7.1 (4.4-9.8) | < 0.001 |
| | Large pupil | 9506 | 188 (2.0%) | 76.0 (72.3-79.6) | 79.4 (76.1-82.7) | 3.4 (-0.8-7.6) | 0.055 |

**Supplementary Table 8. Subgroup analysis for vision-threatening diabetic retinopathy (VTDR) in validation set A.** Formatting as per Supplementary Table 5. Bold indicates there is a >5% decrease in AUC for the DLS in that subgroup compared to the entire cohort.

| Category | Subgroup | No. of cases | | AUC (%) (95% CI) | | | P-value for improvement |
|---|---|---|---|---|---|---|---|
| | | Total | Positive (%) | Baseline characteristics | DLS (external eye image) | Improvement | |
| All | | 26950 | 1042 (3.9%) | 72.7 (71.1-74.2) | 79.4 (78.1-80.8) | 6.8 (4.9-8.7) | < 0.001 |
| Age | ≤50 years old | 10075 | 368 (3.7%) | 75.5 (73.0-78.1) | 80.0 (77.8-82.3) | 4.5 (1.5-7.6) | 0.002 |
| | 50-60 years old | 8937 | 381 (4.3%) | 70.2 (67.5-72.8) | 79.1 (76.8-81.4) | 8.9 (5.7-12.1) | < 0.001 |
| | >60 years old | 8415 | 313 (3.7%) | 71.8 (69.0-74.7) | 75.9 (73.2-78.7) | 4.1 (0.4-7.8) | 0.015 |
| Sex | Male | 12659 | 565 (4.5%) | 70.7 (68.5-72.9) | 77.8 (75.9-79.7) | 7.1 (4.4-9.9) | < 0.001 |
| | Female | 14301 | 514 (3.6%) | 73.4 (71.2-75.6) | 79.3 (77.4-81.3) | 5.9 (3.3-8.6) | < 0.001 |
| Race / ethnicity | Hispanic | 11177 | 457 (4.1%) | 71.4 (69.0-73.8) | 81.1 (79.1-83.0) | 9.6 (6.8-12.5) | < 0.001 |
| | White | 7528 | 211 (2.8%) | 74.7 (71.4-78.0) | 77.9 (74.9-81.0) | 3.2 (-0.9-7.4) | 0.064 |
| | **Black** | **4094** | **176 (4.3%)** | **70.8 (66.9-74.8)** | **74.1 (70.5-77.8)** | **3.3 (-1.7-8.3)** | **0.098** |
| | Asian / Pacific islander | 2882 | 141 (4.9%) | 69.1 (64.5-73.8) | 81.6 (77.9-85.3) | 12.4 (7.3-17.5) | < 0.001 |
| No. years with diabetes | ≤5 | 15059 | 265 (1.8%) | 61.7 (58.3-65.1) | 79.8 (77.1-82.5) | 18.1 (14.0-22.3) | < 0.001 |
| | 5-10 | 7023 | 253 (3.6%) | 56.9 (53.3-60.6) | 78.2 (75.6-80.9) | 21.3 (17.0-25.6) | < 0.001 |
| | **>10** | **6457** | **570 (8.8%)** | **60.8 (58.4-63.2)** | **73.9 (71.8-75.9)** | **13.1 (10.2-15.9)** | **< 0.001** |
| Cataract | **Present** | **656** | **20 (3.0%)** | **68.5 (54.8-82.1)** | **73.8 (62.9-84.6)** | **5.3 (-10.2-20.7)** | **0.251** |
| | Absent | 26413 | 1039 (3.9%) | 72.5 (70.9-74.1) | 79.4 (78.0-80.7) | 6.9 (5.0-8.8) | < 0.001 |
| HbA1c (%) | ≤7 | 8130 | 93 (1.1%) | 69.3 (63.2-75.4) | 77.4 (72.7-82.1) | 8.1 (1.3-14.9) | 0.010 |
| | 7-9 | 7972 | 297 (3.7%) | 70.4 (67.4-73.5) | 79.6 (77.1-82.2) | 9.2 (5.5-12.9) | < 0.001 |
| | >9 | 6665 | 446 (6.7%) | 68.4 (65.9-71.0) | 75.4 (73.1-77.6) | 7.0 (3.8-10.1) | < 0.001 |
| Pupil diameter | Small pupil | 10645 | 473 (4.4%) | 69.8 (67.4-72.1) | 77.8 (75.8-79.8) | 8.0 (5.1-10.9) | < 0.001 |
| | Medium pupil | 17565 | 712 (4.1%) | 71.8 (69.9-73.7) | 78.3 (76.6-80.0) | 6.5 (4.1-8.9) | < 0.001 |
| | Large pupil | 9506 | 241 (2.5%) | 76.8 (73.7-80.0) | 81.4 (78.6-84.2) | 4.5 (0.9-8.2) | 0.008 |

**Supplementary Table 9. Subgroup analysis for cataract in validation set A.** Formatting as per Supplementary Table 5. Bold indicates there is a >5% decrease in AUC for the DLS in that subgroup compared to the entire cohort.

| Category | Subgroup | No. of cases | | AUC (%) (95% CI) | | | |
|---|---|---|---|---|---|---|---|
| | | Total | Positive (%) | Baseline characteristics | DLS (external eye image) | Improvement | P-value for improvement |
| All | | 27415 | 639 (2.3%) | 76.5 (74.7-78.2) | 86.7 (85.3-88.1) | 10.2 (8.6-11.9) | < 0.001 |
| Age | ≤50 years old | 10198 | 56 (0.5%) | 63.8 (56.1-71.4) | 87.1 (81.0-93.3) | 23.4 (13.6-33.1) | < 0.001 |
| | **50-60 years old** | **9093** | **150 (1.6%)** | **62.3 (57.8-66.8)** | **79.5 (75.4-83.5)** | **17.2 (11.7-22.6)** | **< 0.001** |
| | **>60 years old** | **8625** | **448 (5.2%)** | **62.9 (60.4-65.5)** | **80.0 (77.8-82.2)** | **17.0 (14.2-19.9)** | **< 0.001** |
| Sex | Male | 12890 | 247 (1.9%) | 74.1 (71.0-77.1) | 84.6 (82.1-87.1) | 10.6 (7.4-13.7) | < 0.001 |
| | Female | 14525 | 390 (2.7%) | 77.5 (75.5-79.6) | 87.8 (86.1-89.5) | 10.2 (8.2-12.3) | < 0.001 |
| Race / ethnicity | Hispanic | 11393 | 167 (1.5%) | 77.2 (73.7-80.7) | 89.2 (86.8-91.6) | 12.1 (8.5-15.6) | < 0.001 |
| | White | 7621 | 162 (2.1%) | 72.2 (68.1-76.2) | 83.8 (80.4-87.2) | 11.6 (7.6-15.6) | < 0.001 |
| | Black | 4140 | 90 (2.2%) | 76.5 (72.4-80.7) | 83.3 (78.9-87.7) | 6.8 (2.4-11.1) | 0.001 |
| | Asian / Pacific islander | 2949 | 184 (6.2%) | 72.9 (69.7-76.1) | 85.8 (83.1-88.5) | 12.9 (9.5-16.4) | < 0.001 |
| No. years with diabetes | ≤5 | 15204 | 284 (1.9%) | 78.8 (76.1-81.5) | 88.6 (86.5-90.7) | 9.8 (7.1-12.4) | < 0.001 |
| | 5-10 | 7137 | 171 (2.4%) | 72.7 (69.0-76.4) | 85.6 (82.7-88.6) | 12.9 (9.2-16.7) | < 0.001 |
| | >10 | 6677 | 214 (3.2%) | 71.5 (68.3-74.6) | 82.9 (80.2-85.6) | 11.4 (8.0-14.8) | < 0.001 |
| HbA1c (%) | ≤7 | 8214 | 223 (2.7%) | 78.7 (76.0-81.5) | 89.0 (86.8-91.2) | 10.3 (7.7-12.8) | < 0.001 |
| | 7-9 | 8080 | 194 (2.4%) | 76.0 (72.6-79.4) | 87.8 (85.5-90.2) | 11.8 (8.5-15.1) | < 0.001 |
| | >9 | 6825 | 150 (2.2%) | 75.9 (72.5-79.4) | 85.4 (82.4-88.3) | 9.5 (5.5-13.4) | < 0.001 |
| Pupil diameter | **Small pupil** | **11004** | **293 (2.7%)** | **74.1 (71.5-76.7)** | **81.4 (79.1-83.8)** | **7.3 (4.7-10.0)** | **< 0.001** |
| | Medium pupil | 17785 | 447 (2.5%) | 76.2 (74.1-78.3) | 87.4 (85.7-89.1) | 11.2 (9.0-13.4) | < 0.001 |
| | Large pupil | 9554 | 171 (1.8%) | 80.9 (77.7-84.1) | 90.2 (87.5-92.9) | 9.3 (6.4-12.2) | < 0.001 |

**Supplementary Table 10. Hyperparameters that are (A) common across models and (B) specific to each selected model.**

**A**

| Hyperparameter | DLS (external eye image) | Pupil / iris model |
|---|---|---|
| Batch size per TPU core | 8 | 8 |
| Number of TPU cores | 8 | 2 |
| Effective batch size | 64 | 16 |
| Moving average decay rate | 0.9999 | 0.9999 |
| Max training steps | 300000 | 300000 |
| Input image size [pixels] | 587 × 587 | 587 × 587 |
| Random flip | Horizontal and vertical | - |
| Random brightness max delta | 0.1148 | 0.1148 |
| Random hue max delta | 0.0251 | 0.0251 |
| Random saturation range | 0.5597 - 1.2749 | 0.5597 - 1.2749 |
| Random contrast range | 0.9997 - 1.7705 | 0.9997 - 1.7705 |
| Random scale factors | 1.0X (60%), 1.3X (20%), 1.5X (20%) | 1.0X (100%) |
| Dropout rate at final layer | 0.2 | 0.2 |
| Weight decay rate (L2 regularization) | 4.0e-05 | 4.0e-05 |
| Model architecture | Inception V3 | Inception V3 |
| Preinitialization | ImageNet | ImageNet |

**B**

| Hyperparameter | DLS 1 | DLS 2 | DLS 3 | DLS 4 | DLS 5 | Pupil / iris model |
|---|---|---|---|---|---|---|
| Optimizer | Momentum | Momentum | Momentum | Momentum | Momentum | Momentum |
| Initial learning rate | 1.39e-02 | 2.53e-02 | 2.17e-02 | 3.58e-03 | 1.38e-02 | 5.68e-02 |
| Decay steps | 10000 | 20000 | 10000 | 50000 | 50000 | 10000 |
| Decay rate | 0.662 | 0.545 | 0.610 | 0.956 | 0.685 | 0.933 |
| Pre warm-up learning rate | 1.39e-03 | 2.53e-03 | 2.17e-03 | 3.58e-04 | 1.38e-03 | 1.39e-03 |
| Warm up steps | 4000 | 4000 | 4000 | 4000 | 4000 | 4000 |
| Training steps (early stopping) | 52000 | 90000 | 78000 | 80000 | 136000 | 298000 |